\documentclass[aps, prb, superscriptaddress, floatfix, 10pt]{revtex4-2}
\usepackage{appendix}
\usepackage[normalem]{ulem}
\usepackage[utf8]{inputenc}
\usepackage{amsmath, amssymb, graphicx, appendix, comment}
\usepackage{url, booktabs, bm, stackrel, amssymb, xcolor, lineno}
\usepackage[colorlinks=true, citecolor=blue, urlcolor=blue]{hyperref}

\begin{document}

\title{Emergence of chiral \texorpdfstring{\(p\)-wave}{p-wave} and \texorpdfstring{\(d\)-wave}{d-wave} states in \texorpdfstring{\(g\)-wave}{g-wave} altermagnets}

\date{\today}

\author{Tilen \v{C}ade\v{z}}
\affiliation{Asia Pacific Center for Theoretical Physics, Pohang, Gyeongbuk, 37673, Republic of Korea}

\author{Abraham Nathan Sunanta}
\affiliation{Department of Physics, Faculty of Mathematics and Natural Sciences, Universitas Indonesia, Depok 16424, Indonesia}

\author{Kyoung-Min Kim}
\email{kyoungmin.kim@apctp.org}
\affiliation{Asia Pacific Center for Theoretical Physics, Pohang, Gyeongbuk, 37673, Republic of Korea}
\affiliation{Department of Physics, Pohang University of Science and Technology, Pohang, Gyeongbuk 37673, Korea}

\begin{abstract}
    Altermagnets emerge as a novel platform for realizing unconventional superconductivity through their exotic momentum-dependent spin-splitting of electronic band structures. Recent experiments have uncovered a novel form of altermagnetism with distinctive \(g\)-wave symmetry in CrSb. However, the potential for unconventional superconductivity arising from \(g\)-wave altermagnetism in such systems remains largely unexplored. In this study, we discover the emergence of chiral superconducting states in three-dimensional \(g\)-wave altermagnetic metals. Through systematic self-consistent mean-field analysis on the extended attractive Hubbard model combined with \(g\)-wave altermagnetic exchange fields in a three-dimensional hexagonal lattice, as observed in CrSb, we find that the altermagnetic spin splitting of Fermi surfaces favors chiral \(p\)-wave states as the dominant pairing channel under strong altermagnetic fields and high electron densities, while chiral \(d\)-wave states become predominant under weak altermagnetic fields and intermediate electron densities. Conversely, at weak altermagnetic fields and typical electron densities, non-chiral \(s\)-, extended \(s\)-, or \(f\)-wave states become stabilized. We also showcase the possible experimental detection using the quasiparticle energy dispersions and the density of states to distinguish different pairing symmetries. These findings underscore the potential of \(g\)-wave altermagnets to host sought-after chiral and gapless superconductivity.    
\end{abstract}

\keywords{Chiral superconductivity, $g$-wave altermagnets, CrSb, BdG analysis}

\maketitle

\tableofcontents

\section{Introduction}

Altermagnetism is a recently identified class of magnetism characterized by a unique combination of antiferromagnetic-like spin arrangements and crystallographic rotational symmetries, distinguishing it from conventional ferromagnetism or antiferromagnetism \cite{https://doi.org/10.1002/adfm.202409327, PhysRevX.12.040501, Cheong2025, Song2025}. Due to this distinctive physical attribute, the electronic band structures of altermagnets exhibit momentum-dependent spin splitting, often referred to as ``altermagnetic spin splitting," even without conventional mechanisms such as net magnetization or spin-orbit coupling \cite{PhysRevX.12.031042}. An important perspective is how the altermagnetic spin splitting influences various superconducting phenomena. For example, previous studies have shown that \(d\)-wave altermagnetic spin splitting can promote unconventional superconducting states, such as topological chiral \(p\)-wave states \cite{PhysRevB.108.184505, PhysRevB.111.054501, 4318-ttvf}, gapless superconductivity with a Bogoliubov Fermi surface (BFS) \cite{PhysRevB.109.L201404, PhysRevB.111.054501}, and finite-momentum Cooper pairing \cite{PhysRevB.110.L060508, Zhang2024, PhysRevB.110.L140506, PhysRevB.109.L220505, PhysRevB.111.054501}. Moreover, the impact of \(d\)-wave altermagnetic spin splitting on the Josephson junction effect \cite{PhysRevLett.131.076003, PhysRevLett.134.026001, PhysRevB.109.024517, PhysRevB.109.134511}, Majorana edge modes \cite{PhysRevB.108.205410, PhysRevB.108.184505, PhysRevLett.133.106601}, Andreev reflection \cite{PhysRevB.108.L060508, PhysRevB.108.054511}, thermoelectric effects \cite{PhysRevB.110.094508, PhysRevB.111.224503}, and the superconducting diode effect \cite{PhysRevB.110.024503, PhysRevB.110.014518} has also been explored.

Despite substantial previous focus on \(d\)-wave altermagnets \cite{PhysRevB.111.054501, PhysRevB.110.205120, PhysRevB.108.224421, PhysRevB.109.L201404, PhysRevB.111.054501, PhysRevB.110.L060508, Zhang2024, PhysRevB.110.L140506, PhysRevB.109.L220505, PhysRevB.111.054501, PhysRevLett.131.076003, PhysRevLett.134.026001, PhysRevB.109.024517, PhysRevB.109.134511, PhysRevB.108.205410, PhysRevB.108.184505, PhysRevLett.133.106601, PhysRevB.108.L060508, PhysRevB.108.054511, PhysRevB.110.094508, PhysRevB.111.224503, PhysRevB.110.024503, PhysRevB.110.014518, cv8s-tk4c, dlpb-gfct, b7rh-v7nq, PhysRevB.110.205120, PhysRevB.108.224421, 4318-ttvf}, the exploration of superconductivity in altermagnets extends beyond this specific form, encompassing broader higher-order even-parity wave characteristics \cite{PhysRevX.12.031042}. Particular noteworthy are \(g\)-wave altermagnets that have been theoretically predicted \cite{PhysRevB.107.L100418, PhysRevLett.134.086701, PhysRevX.12.031042} and observed in recent experiments \cite{PhysRevLett.133.206401, Reimers2024, Yang2025, Krempasky2024, PhysRevLett.132.036702, https://doi.org/10.1002/adma.202508977}. Key features of \(g\)-wave altermagnetic metals in a three-dimensional hexagonal lattice, such as CrSb, include a large spin splitting of Fermi surfaces for spin-up and spin-down electrons and the presence of sixfold nodal planes where the energy gap closes \cite{PhysRevLett.133.206401, Reimers2024, Yang2025}. A central question is how these unique spin-split Fermi surfaces influence the dominant superconducting instability, potentially transforming it from conventional \(s\)-wave pairing into other intriguing unconventional pairing states. Of particular interest are chiral superconducting states that carry non-zero angular momentum and often feature topologically protected Majorana edge modes \cite{Kallin_2016}. Despite numerous proposals and experimental signatures suggestive of their existence \cite{Li2016, PhysRevB.108.184505, PhysRevB.111.054501, 4318-ttvf, PhysRevB.89.144501, PhysRevB.94.115105, PhysRevLett.121.217001, Jiao2020, Biswas2021, Ghosh2021, 10.21468/SciPostPhys.11.1.017, Yin2022, Ming2023, Deng2024, Han2025}, the realization of chiral superconductivity still remains elusive and warrants further investigation. Achieving such states would offer a crucial platform for topological physics studies \cite{Sato_2017} and enable fault-tolerant quantum computing via Majorana modes \cite{RevModPhys.80.1083}.

In this paper, we investigate the potential realization of chiral superconductivity in \(g\)-wave altermagnetic metals through a Bogoliubov-de Gennes (BdG) analysis of an extended attractive Hubbard model relevant to three-dimensional \(g\)-wave altermagnetic metals such as CrSb. We examine all possible intralayer pairing channels in hexagonal lattice systems, including \(s\)-wave, extended \(s\)-wave, \(d+id\)-wave, \(f\)-wave, and \(p+ip\)-wave pairing symmetries, in addition to interlayer pairing channels. By numerically solving self-consistent gap equations for these pairing states and comparing their condensation energies, we demonstrate that chiral \(d+id\)-wave or \(p+ip\)-wave pairing state is energetically favored over non-chiral pairing states due to the altered phase space available for Cooper pairing resulting from the altermagnetic spin splitting through the formation of BFSs \cite{PhysRevLett.118.127001}. Finally, we investigate the BdG quasiparticle energy dispersions and the corresponding density of states, which could serve as an experimental probe to distinguish between different pairing symmetries.

\begin{figure}[t!]
    \centering
    \includegraphics[width=1\columnwidth]{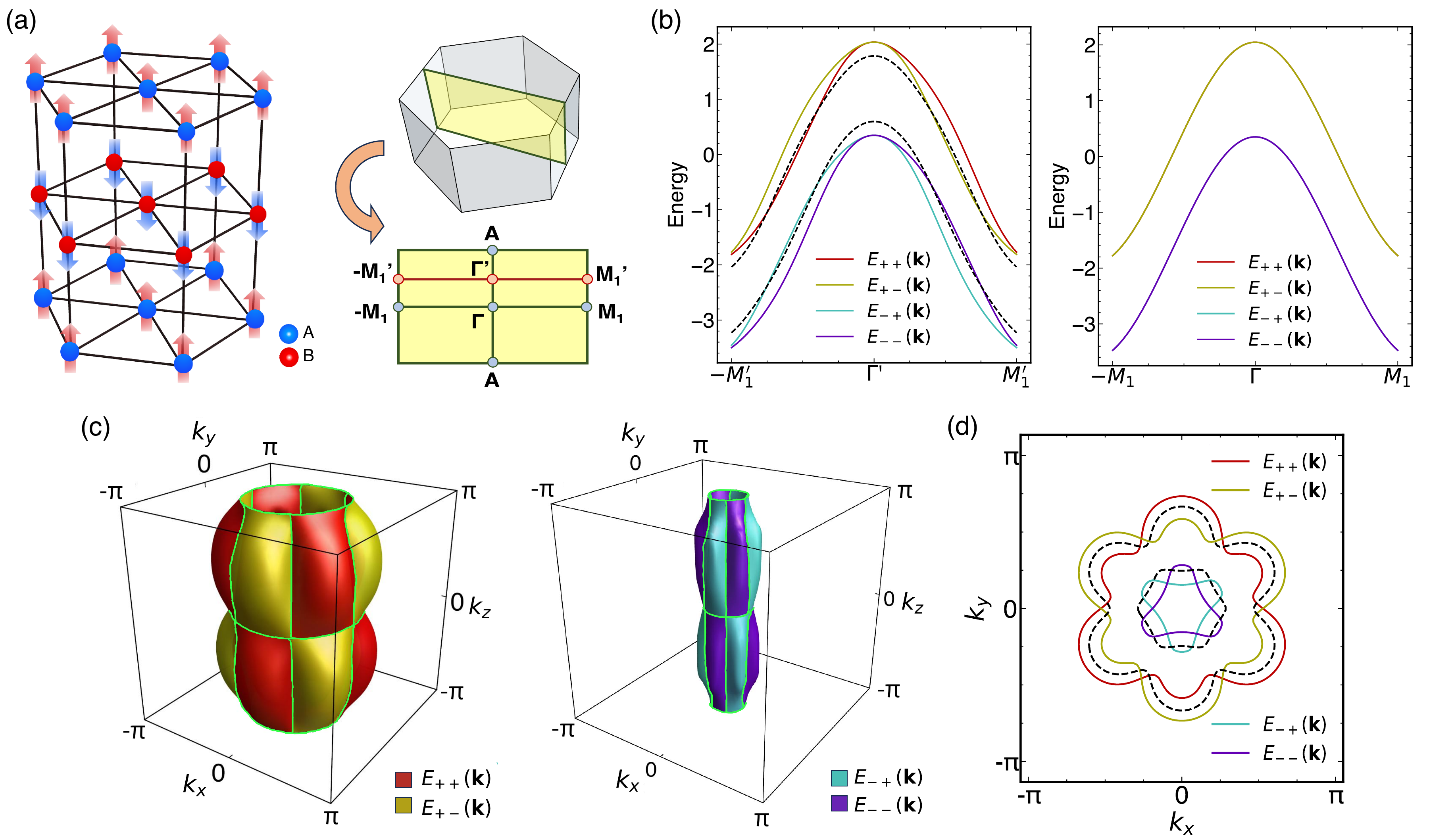}
\caption{ 
    \textbf{Altermagnetic spin splitting in electronic band structures.}
    (a) Illustration of a three-dimensional hexagonal lattice (left) and its Brillouin zone (BZ; right). In the left panel, blue and red spheres represent the A and B sublattice layers, respectively. Red and blue arrows indicate magnetic moments oriented in opposite directions on these layers. In the right panel, high-symmetry points \(\Gamma\), \(\Gamma'\), \(M_{1}\), \(M_{1}'\), and \(A\) are marked.
    (b) Electronic band structures \(E_{\alpha\beta}(\bm{k})\) along high-symmetry lines. The four bands correspond to combinations of band index \(\alpha=\pm\) and spin index \(\beta=\pm\). In the left panel, for each \(\alpha\), the spin-up (\(\beta=+\)) and spin-down (\(\beta=-\)) bands exhibit \(g\)-wave altermagnetic spin splitting along the high-symmetry line \(-M_{1}'\)–\(\Gamma\)–\(M_{1}'\). Conversely, in the right panel, no spin splitting is observed along \(-M_{1}\)–\(\Gamma\)–\(M_{1}\).
    (c) Spin-split Fermi surfaces derived from (left) the upper band (\(\alpha=+\)) and (right) the lower band (\(\alpha=-\)). In each panel, green lines highlight characteristic nodal planes.
    (d) Two-dimensional Fermi surface projections onto the \(k_x\)-\(k_y\) plane at \(k_z=\pi/2\). }
    \label{fig:model}
\end{figure}

\section{Result}

\subsection{Extended attractive Hubbard model}

We consider a three-dimensional hexagonal lattice with two sublattice layers, belonging to space group 194, with point group D\textsubscript{6h} [Fig.~\ref{fig:model}(a)]. To describe the electronic bands exhibiting \(g\)-wave altermagnetic spin splitting in our system, we adopt the following  tight-binding model derived for CrSb \cite{PhysRevB.110.144412}:
\begin{equation}
    \hat{H}_0 = \sum_{\bm{k}} \sum_{s,s'}\sum_{\sigma, \sigma'} c_{\bm{k}s\sigma}^\dagger T_{ss'}^{\sigma\sigma'}(\bm{k}) c_{\bm{k}s'\sigma'}. \label{eq:H_0}
\end{equation}
In this expression, the operator $c_{\bm{k}s\sigma}^\dagger$ creates an electron with momentum $\bm{k}$ in sublattice layer $s$ with spin $\sigma$. The summations over $s,s'=\{a,b\}$ and $\sigma,\sigma'=\{\uparrow,\downarrow\}$ are implied throughout this paper. The matrix form of the Bloch Hamiltonian $T_{ss'}^{\sigma\sigma'}(\bm{k})$ is given by:
\begin{equation}
    T(\bm{k}) = \epsilon_{0}(\bm{k}) \tau_0\sigma_0 + t_{x}(\bm{k})\tau_x\sigma_0 + t_{z}(\bm{k})\tau_z \sigma_0  + \tau_z \bm{J}\cdot\bm{\sigma}, \label{eq:kinetic_term}
\end{equation}
where the first term represents the average kinetic energy between the two sublattice layers, the second term describes hopping between the two sublattice layers within the same unit cell, the third term captures the local point group symmetry-breaking effect, reflecting the kinetic energy difference between the layers, and the fourth term represents the exchange field originating from the ordered magnetic moments in the Néel antiferromagnetic state. We choose the direction of $\bm{J}$ as $\bm{J} = J\hat{z}$ without loss of generality. The matrices $\tau_{x,y,z}$ are the Pauli matrices acting on the sublattice degree of freedom, while $\bm{\sigma}=(\sigma_x,\sigma_y,\sigma_z)$ are the Pauli matrices for spin. The $2 \times 2$ identity matrices $\tau_0$ and $\sigma_0$ act in the sublattice and spin degrees of freedom, respectively. Considering the space group symmetry, the momentum-dependent functions in the first three terms are determined as follows \cite{PhysRevB.110.144412}:
\begin{equation}
\begin{aligned}
    \epsilon_{0}(\bm{k}) & = t_1 \cos{k_x} + 2 t_1\cos{\frac{k_x}{2}}\cos{\frac{\sqrt{3}k_y}{2}} + t_2 \cos{k_z} - \mu, \\
    t_{x}(\bm{k}) & = t_3 \cos{\frac{k_z}{2}}, \\
    t_{z}(\bm{k}) & = t_4\sin{k_z}f_{y}(\bm{k}) (f_{y}(\bm{k})^2 - 3f_{x}(\bm{k})^2),
\end{aligned}
\end{equation}
with $f_{x}(\bm{k})$ and $f_{y}(\bm{k})$ defined as
\begin{equation}
\begin{aligned}
    f_{x}(\bm{k}) & = \sin{k_x} + \sin{\frac{k_x}{2}}\cos{\frac{\sqrt{3}k_y}{2}}, \\
    f_{y}(\bm{k}) & = \sqrt{3} \cos{\frac{k_x}{2}}\sin{\frac{\sqrt{3}k_y}{2}}.
\end{aligned}
\end{equation}
In the function \(\epsilon_{0}(\bm{k})\), the parameter $t_1$ denotes the hopping amplitude between nearest-neighbor sites within a sublattice layer, $t_2$ accounts for hopping to a next-nearest neighbor layer in a different unit cell, and $\mu$ is the chemical potential. In our analysis, we employ the following hopping parameters:
\begin{equation}
    t_1 = 1, \quad t_2 = 0.2, \quad t_3 = 0.6, \quad t_4 = 0.3. \label{eq:hopping_params}
\end{equation}
With these parameters along with \(\mu=2\) and \(J=0.6\), the electronic bands display significant \(g\)-wave spin-splitting [Fig.~\ref{fig:model}(b)], which closely match those of CrSb \cite{Reimers2024}. All physical constants are measured in meV units.

The four energy bands, obtained by diagonalizing the Bloch Hamiltonian in Eq.~\eqref{eq:kinetic_term}, are expressed as:
\begin{equation}
    E_{\alpha\beta}(\bm{k}) = \epsilon_{0}(\bm{k}) + \alpha \sqrt{t_{x}(\bm{k})^2 + (t_{z}(\bm{k}) + \beta J)^2}, \label{eq:E_k}
\end{equation}
where $\alpha = +$ ($\alpha = -$) denotes the upper (lower) band, while $\beta = +$ ($\beta = -$) corresponds to the spin-up (spin-down) state. In these bands, the combined effects of the Néel order magnetization (\(J\)) and the local point group symmetry-breaking effect (\(t_{z}(\bm{k})\)) causes momentum-dependent energy splitting in the spin-up and spin-down bands for a given $\alpha$ \cite{PhysRevB.110.144412}, as illustrated in Fig.~\ref{fig:model}(b). The Fermi surfaces feature spin-degenerate nodal lines parallel to the \(k_z\) axis, and an additional nodal lines in the \(k_z = 0\) plane [green lines in Fig.~\ref{fig:model}(c)], characteristic of $g$-wave altermagnetism in a three-dimensional hexagonal lattice \cite{PhysRevX.12.031042}. Furthermore, this \(g\)-wave altermagnetic spin splitting breaks both sixfold rotational and time-reversal symmetry—defined respectively as $\mathcal{C}_{6z}$: $(k_x,k_y,k_z)\rightarrow\left(\frac{k_x-\sqrt{3}k_y}{2},\frac{\sqrt{3}k_x+k_y}{2},k_z\right)$ and $\mathcal{T}$: $\bm{k}\rightarrow-\bm{k},~\beta\rightarrow -\beta$—while respecting their combined symmetry \(\mathcal{T}\mathcal{C}_{6z}\) [Fig.~\ref{fig:model}(d)].

We additionally incorporate an interaction Hamiltonian \(\hat{H}_\textrm{int}\), which describes phonon-mediated attractive interactions between electrons, into our model. The interaction Hamiltonian is given by:
\begin{equation}
\begin{aligned}
    \hat{H}_\textrm{int} = & \; -U\sum_{i} \sum_{s} n_{is\uparrow}n_{is\downarrow} -V_1 \sum_{\langle i, j\rangle_{\parallel}} \sum_{s} \sum_{\sigma,\sigma'} n_{is\sigma} n_{js\sigma'} \\
    & -V_{2}\sum_{i} \sum_{\sigma,\sigma'}(n_{ia\sigma}n_{ib\sigma'} + n_{ib\sigma}n_{i+za\sigma'} ), \label{eq:H_int_real_space}
\end{aligned}
\end{equation}
where the first term represents the onsite interaction, the second term accounts for the nearest neighbor interaction within the same layer ($\langle i,j\rangle_\parallel$ signifies nearest neighbor sites within the same layer), and the third term indicates the interaction between two adjacent layers from different sublattices. The momentum-space representation of \(\hat{H}_\textrm{int}\) reads
\begin{equation}
    \hat{H}_\textrm{int} = - \sum_{\bm{k},\bm{p},\bm{q}} \sum_{s,s'}\sum_{\sigma,\sigma'} \frac{V_{ss'}^{\sigma\sigma'}(\bm{q})}{N} c_{\bm{k}+\bm{q}s\sigma}^\dagger c_{\bm{k}s\sigma} c_{\bm{p}-\bm{q}s'\sigma'}^\dagger c_{\bm{p}s'\sigma'}, \label{eq:H_int_momentum_space}
\end{equation}
where the matrix form of the interaction function $V_{ss'}^{\sigma\sigma'}(\bm{q})$ is given by:
\begin{equation}
    V(\bm{q}) = \frac{1}{2}U\tau_0 \sigma_{x} + V_1 \tau_0 \sigma_{0x} \Bigg[\cos{q_x} + 2\cos\frac{q_x}{2}\cos\frac{\sqrt{3}q_y}{2}\Bigg] + \frac{1}{2}V_2\tau_x \sigma_{0x}\cos\frac{q_z}{2},
\end{equation}
with $\sigma_{0x}=\sigma_0+\sigma_x$ and $N$ denoting the total number of unit cells. Although, in principle, \(U\) and \(V_1\) could differ between sublattices, such sublattice dependence is neglected in this study. The total Hamiltonian is given by:
\begin{equation}
    \hat{H} = \hat{H}_0 + \hat{H}_\textrm{int}. \label{eq:H}
\end{equation}

\subsection{Mean-field analysis using BdG formalism} \label{secIIB}

For our mean-field analysis, we employ the BdG Hamiltonian \(\hat{H}_\textrm{BdG} = \hat{H}_0 + \hat{H}_\textrm{p}\), where the pairing Hamiltonian \(\hat{H}_\textrm{p}\) approximates \(\hat{H}_\textrm{int}\) at the mean-field level. In this approach, the onsite interaction gives rise to spin-singlet pairing while the nearest neighbor and interlayer interactions can mediate either spin-singlet or spin-triplet pairing. The absence of spin-orbit coupling ensures that spin-singlet and spin-triplet states remain decoupled, allowing us to consider them separately. To describe spin-singlet states, we utilize the following BdG Hamiltonian (derivation is presented in Appendix \ref{app:H_BdG}):
\begin{equation}
\label{eq:H_BdG_singlet}
    \hat{H}_\textrm{BdG}^\textrm{singlet} = \hat{H}_0 + \sum_{\bm{k}} \sum_{s,s'} \bigg[ \Delta_{ss'}^{\uparrow\downarrow}(\bm{k}) c_{\bm{k}s\uparrow}^\dagger c_{-\bm{k}s'\downarrow}^\dagger + \textrm{h.c.} \bigg] + E_\Delta, 
\end{equation}
where the pairing gap function \(\Delta_{ss'}^{\uparrow\downarrow}(\bm{k})\) is expressed as:
\begin{equation}
\label{eq:Delta_k_singlet}
    \Delta_{ss'}^{\uparrow\downarrow}(\bm{k}) = \sum_{\eta=s,es,d+id,d-id,s_z}2\Delta_{ss';\eta}^{\uparrow\downarrow} g_{\eta}(\bm{k}).
\end{equation}
The pairing amplitudes \(\Delta_{ss;\eta}^{\uparrow\downarrow}\) for \(\eta=\{s,es,d+id,d-id,s_z\}\) are obtained via self-consistent equations:
\begin{equation}
\begin{aligned}
\label{eq:self_cons_eq_singlet_app}
    \Delta_{ss;s}^{\uparrow\downarrow} = & -\frac{U}{2N}\sum_{\bm{k'}} g_{s}^*(\bm{k}') \langle c_{-\bm{k}'s\downarrow} c_{\bm{k}'s\uparrow} \rangle, \\
    \Delta_{ss;es}^{\uparrow\downarrow} = & -\frac{V_{1}}{3N}\sum_{\bm{k'}} g_{es}^*(\bm{k}') \langle c_{-\bm{k}'s\downarrow} c_{\bm{k}'s\uparrow} \rangle, \\
    \Delta_{ss;d\pm id}^{\uparrow\downarrow} = & -\frac{V_{1}}{3N}\sum_{\bm{k'}} g_{d\pm id}^*(\bm{k}') \langle c_{-\bm{k}'s\downarrow} c_{\bm{k}'s\uparrow} \rangle, \\
    \Delta_{s\bar{s};s_{z}}^{\uparrow\downarrow} = & -\frac{V_{2}}{2N}\sum_{\bm{k'}} g_{s_{z}}^*(\bm{k}') \langle c_{-\bm{k}'\bar{s}\downarrow} c_{\bm{k}'s\uparrow} \rangle,
\end{aligned}
\end{equation}
where $\bar{s}=b$ if $s=a$ and $\bar{s}=a$ if $s=b$. The form factors \(g_{\eta}(\bm{k})\) for \(\eta=\{s,es,d+id,d-id,s_z\}\) have even parity under inversion \(\bm{k} \to -\bm{k}\):
\begin{equation}
\begin{aligned}
\label{eq:g_k_singlet}
    g_s(\bm{k}) &= 1, \\
    g_{es}(\bm{k}) &= \cos k_x + 2 \cos \frac{k_x}{2} \cos \frac{\sqrt{3}k_y}{2}, \\
    g_{d\pm id}(\bm{k}) &= \cos k_x - \cos \frac{k_x}{2} \cos \frac{\sqrt{3}k_y}{2} \pm i \sqrt{3} \sin \frac{k_x}{2} \sin \frac{\sqrt{3}k_y}{2}, \\
    g_{s_z}(\bm{k}) &= \cos \frac{k_z}{2}.
\end{aligned}
\end{equation}
The pairing energy cost \(E_\Delta\) is given by \(E_\Delta = N \sum_{s} \Big( \frac{4\lvert\Delta_{ss;s}^{\uparrow\downarrow}\rvert^2}{U} + \frac{6|\Delta_{ss;es}^{\uparrow\downarrow}|^2}{V_1} + \frac{6|\Delta_{ss;d+id}^{\uparrow\downarrow}|^2}{V_1} + \frac{6|\Delta_{ss;d-id}^{\uparrow\downarrow}|^2}{V_1} + \frac{4|\Delta_{s\bar{s};s_{z}}^{\uparrow\downarrow}|^2}{V_2} \Big)\).

Similarly, spin-triplet states are described by the BdG Hamiltonian (derivation is presented in Appendix \ref{app:H_BdG}):
\begin{equation}
\label{eq:H_BdG_triplet}
    \hat{H}_\textrm{BdG}^\textrm{triplet} = \hat{H}_0 + \sum_{\bm{k}} \sum_{s,s'} \sum_{\sigma} \bigg[ \Delta_{ss'}^{\sigma\sigma}(\bm{k}) c_{\bm{k}s\sigma}^\dagger c_{-\bm{k}s'\sigma}^\dagger + \textrm{h.c.} \bigg] + E_\Delta, 
\end{equation}
where the pairing gap function \(\Delta_{ss'}^{\sigma\sigma}(\bm{k})\) for \(\sigma=\uparrow,\downarrow\) is expressed as:
\begin{equation}
\label{eq:Delta_k_triplet}
    \Delta_{ss'}^{\sigma\sigma}(\bm{k}) = \sum_{\eta=f,p+ip,p-ip,p_z}2\Delta_{ss';\eta}^{\sigma\sigma} g_{\eta}(\bm{k}).
\end{equation}
The pairing amplitudes \(\Delta_{ss';\eta}^{\sigma\sigma}\) for \(\eta=\{f,p+ip,p-ip,p_z\}\) are obtained via:
\begin{equation}
\begin{aligned}
\label{eq:self_cons_eq_triplet_app}
    \Delta_{ss;f}^{\sigma\sigma} = & -\frac{V_{1}}{3N}\sum_{\bm{k'}} g_{f}^*(\bm{k}') \langle c_{-\bm{k}'s\sigma} c_{\bm{k}'s\sigma} \rangle, \\
    \Delta_{ss;p\pm ip}^{\sigma\sigma} = & -\frac{V_{1}}{3N}\sum_{\bm{k'}} g_{p\pm ip}^*(\bm{k}') \langle c_{-\bm{k}'s\sigma} c_{\bm{k}'s\sigma} \rangle, \\
    \Delta_{s\bar{s};p_{z}}^{\sigma\sigma} = & -\frac{V_{2}}{2N}\sum_{\bm{k'}} g_{p_{z}}^*(\bm{k}') \langle c_{-\bm{k}'\bar{s}\sigma} c_{\bm{k}'s\sigma} \rangle.
\end{aligned}
\end{equation} 
The form factors \(g_{\eta}(\bm{k})\) for \(\eta=\{f,p+ip,p-ip,p_z\}\) have odd parity:
\begin{equation}
\begin{aligned}
\label{eq:g_k_triplet}
    g_{f}(\bm{k}) &= \sin k_x - 2 \sin \frac{k_x}{2} \cos \frac{\sqrt{3}k_y}{2}, \\
    g_{p\pm ip}(\bm{k}) &= \sin k_x + \sin \frac{k_x}{2} \cos \frac{\sqrt{3}k_y}{2} \pm i \sqrt{3} \cos \frac{k_x}{2} \sin \frac{\sqrt{3}k_y}{2}, \\
    g_{p_z}(\bm{k}) &= \sin \frac{k_z}{2}.
\end{aligned}
\end{equation}
The pairing energy cost \(E_\Delta\) is given by \(E_\Delta = N \sum_{s} \sum_{\sigma} \Big( \frac{3|\Delta_{ss;f}^{\sigma\sigma}|^2}{V_1} + \frac{3|\Delta_{ss;p+ip}^{\sigma\sigma}|^2}{V_1} + \frac{3|\Delta_{ss;p-ip}^{\sigma\sigma}|^2}{V_1} + \frac{2|\Delta_{s\bar{s};p_{z}}^{\sigma\sigma}|^2}{V_2} \Big)\).

\begin{figure}[t!]
    \centering
    \includegraphics[width=.8\columnwidth]{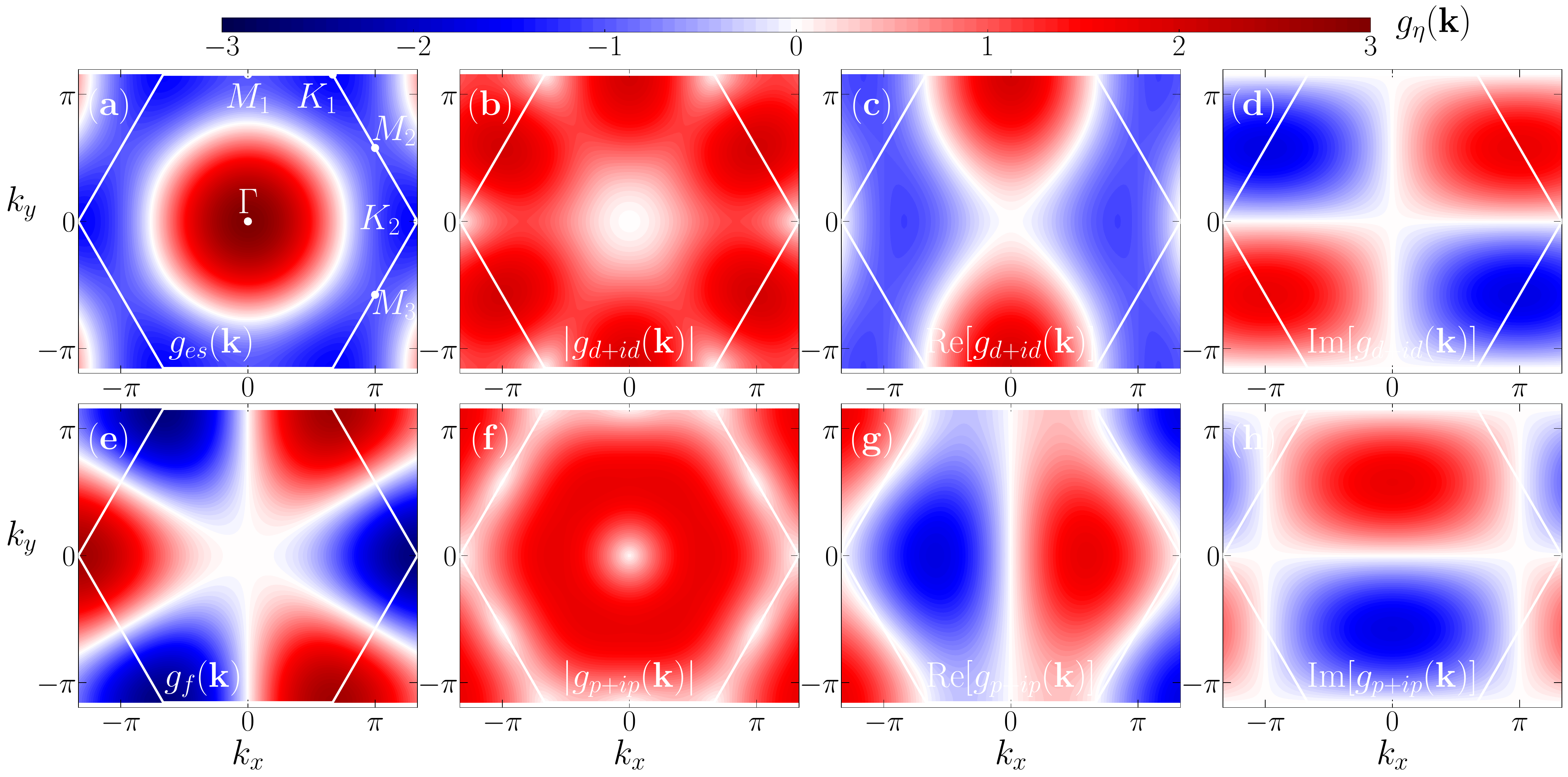}
    \caption{
    \textbf{Form factors of superconducting states.} Each panel depicts: (a) \( g_{es}(\bm{k}) \), (b) \( |g_{d+id}(\bm{k})| \), (c) \( \textrm{Re}[g_{d+id}(\bm{k})] \), (d) \( \textrm{Im}[g_{d+id}(\bm{k})] \), (e) \( g_{f}(\bm{k}) \), (f) \( |g_{p+ip}(\bm{k})| \), (g) \( \textrm{Re}[g_{p+ip}(\bm{k})] \), and (h) \( \textrm{Im}[g_{p+ip}(\bm{k})] \). The functions \( g_{es}(\bm{k}) \), \( g_{d+id}(\bm{k}) \), \( g_{f}(\bm{k}) \), and \( g_{p+ip}(\bm{k}) \) are the form factors associated with the \(es\)-wave, chiral \(d\)-wave, \(f\)-wave, and chiral \(p\)-wave states, respectively.  In each panel, the white solid lines indicate the boundary of the first BZ. In panel (a), K\textsubscript{1}, K\textsubscript{2}, M\textsubscript{1}, M\textsubscript{2}, and M\textsubscript{3} mark the high-symmetry points. 
    In each panel, the \(x\)-axis represents the \(k_x\) value, while the \(y\)-axis indicates the \(k_y\) value.
    }
    \label{fig2:harmonics}
\end{figure}

In our analysis for chiral pairing waves, we focus on the \(d+id\)- and \(p+ip\)-wave pairings, neglecting their complex counterparts—the \(d-id\)- and \(p-ip\)-wave pairings—by setting \(\Delta_{ss;d-id}^{\uparrow\downarrow} = 0\) and \(\Delta_{ss;p-ip}^{\sigma\sigma} = 0\). This simplification is justified because simultaneously considering both \(d+id\)-wave and \(d-id\)-wave pairings (or both \(p+ip\)-wave and \(p-ip\)-wave pairings) tends to increase the total energy of the system, making such configurations energetically less favorable \cite{PhysRevB.108.054511}. This can be understood by examining the pairing gap function: \(\Delta_{ss}^{\uparrow\downarrow}(\bm{k}) = \Delta_0 g_{d+id}(\bm{k}) + \Delta_0 g_{d-id}(\bm{k}) \), whose nodal line structures render it energetically unfavorable compared to \(\Delta_{ss}^{\uparrow\downarrow}(\bm{k}) = \Delta_0 g_{d+id}(\bm{k})\), which features only nodal points. It is also important to note that the \(d-id\)- and \(p-ip\)-wave pairings are degenerate with the \(d+id\)- and \(p+ip\)-wave pairings, respectively. This degeneracy corresponds to the existence of two degenerate chiral \(d\)-wave states: the \(d+id\)-wave state, characterized by \(\Delta_{ss}^{\uparrow\downarrow}(\bm{k}) = \Delta_{ss;d+id}^{\uparrow\downarrow} g_{d+id}(\bm{k})\), and the \(d-id\)-wave state, characterized by \(\Delta_{ss}^{\uparrow\downarrow}(\bm{k}) = \Delta_{ss;d-id}^{\uparrow\downarrow} g_{d-id}(\bm{k})\). For the \(p\)-wave scenario, considering both spin sectors yields four possible states \cite{PhysRevB.108.184505, PhysRevB.108.054511}: (i) the chiral \(p+ip\) state with \(\Delta_{ss}^{\uparrow\uparrow}(\bm{k}) = \Delta_{ss;p+ip}^{\uparrow\uparrow} g_{p+ip}(\bm{k})\) and \(\Delta_{ss}^{\downarrow\downarrow}(\bm{k}) = \Delta_{ss;p+ip}^{\downarrow\downarrow} g_{p+ip}(\bm{k})\); (ii) the chiral \(p-ip\) state with \(\Delta_{ss}^{\uparrow\uparrow}(\bm{k}) = \Delta_{ss;p-ip}^{\uparrow\uparrow} g_{p-ip}(\bm{k})\) and \(\Delta_{ss}^{\downarrow\downarrow}(\bm{k}) = \Delta_{ss;p-ip}^{\downarrow\downarrow} g_{p-ip}(\bm{k})\); (iii) the helical \(p+ip\) state with \(\Delta_{ss}^{\uparrow\uparrow}(\bm{k}) = \Delta_{ss;p+ip}^{\uparrow\uparrow} g_{p+ip}(\bm{k})\) and \(\Delta_{ss}^{\downarrow\downarrow}(\bm{k}) = \Delta_{ss;p-ip}^{\downarrow\downarrow} g_{p-ip}(\bm{k})\); and (iv) the helical \(p-ip\) state with \(\Delta_{ss}^{\uparrow\uparrow}(\bm{k}) = \Delta_{ss;p-ip}^{\uparrow\uparrow} g_{p-ip}(\bm{k})\) and \(\Delta_{ss}^{\downarrow\downarrow}(\bm{k}) = \Delta_{ss;p+ip}^{\downarrow\downarrow} g_{p+ip}(\bm{k})\). All four states are energetically degenerate in our model with inversion symmetry.

We define a superconducting state as a ``pure state'' if its pairing gap function—either given in Eq.~\eqref{eq:Delta_k_singlet} for a spin singlet pairing or in Eq.~\eqref{eq:Delta_k_triplet} for a spin triplet pairing—contains only a single harmonic component (e.g., \(\Delta_{ss}^{\uparrow\downarrow}(\bm{k}) = \Delta_{ss;d+id}^{\uparrow\downarrow} g_{d+id}(\bm{k})\)). Conversely, a superconducting state with a pairing gap function that includes multiple harmonic components is termed a ``mixed state." A pure state incorporates multiple pairing amplitude components due to different sublattices or spin components. A mixed state is further complicated by multiple harmonic components. However, it is crucial to note that not all of these pairing amplitude components are independent due to symmetry constraints. Specifically, in a spin singlet state, pairing wave components from the A and B sublattices are identical: \(\Delta_{aa;\eta}^{\uparrow\downarrow} = \Delta_{bb;\eta}^{\uparrow\downarrow}\) for each harmonic component \(\eta=s,es,d+id\), as \(E_{\alpha\beta}(\bm{k})\) remains the same under the interchange of A \(\leftrightarrow\) B. Consequently, a general spin singlet state can have three independent pairing amplitude components: \(\Delta_{aa;\eta}^{\uparrow\downarrow}\) for \(\eta=s,es,d+id\) (or equivalently \(\Delta_{bb;\eta}^{\uparrow\downarrow}\)). In a spin triplet state, pairing wave components from the spin-up and spin-down sectors are related via: \(\Delta_{aa;\eta}^{\uparrow\uparrow}=\Delta_{bb;\eta}^{\downarrow\downarrow}\) and \(\Delta_{bb;\eta}^{\uparrow\uparrow} = \Delta_{aa;\eta}^{\downarrow\downarrow} \) for \(\eta=f,p+ip\), reflecting the Néel-type staggered configuration between the A and B sublattices. Consequently, a general spin triplet state can have four independent pairing amplitude components: \(\Delta_{aa;\eta}^{\uparrow\uparrow}\) and \(\Delta_{bb;\eta}^{\uparrow\uparrow}\) for \(\eta=f,p+ip\) (or equivalently \(\Delta_{aa;\eta}^{\downarrow\downarrow}\) and \(\Delta_{bb;\eta}^{\downarrow\downarrow}\)).

\subsection{Emergence of chiral \texorpdfstring{\(d\)-wave}{d-wave} and \texorpdfstring{\(p\)-wave}{p-wave} phases} \label{secIIC}

\begin{figure*}[t!]
    \centering
    \includegraphics[width=.99\columnwidth]{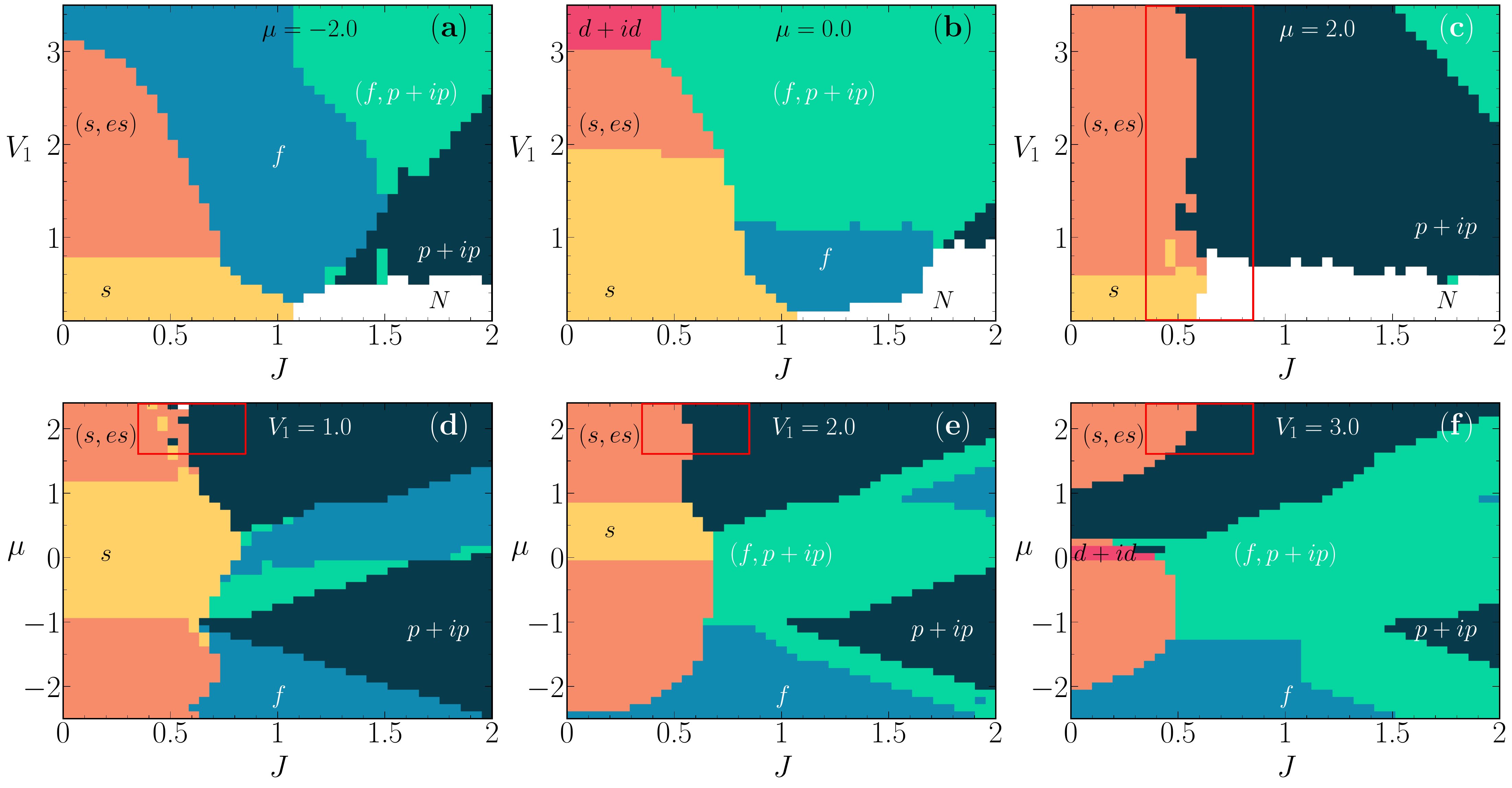}
    \caption{
    \textbf{Superconducting phase diagrams.}
    (a)–(c) Zero temperature phase diagrams as a function of \(J\) and \(V_1\) for three different chemical potential \(\mu\) values: (a) \(\mu=-2\), (b) \(\mu=0\), and (c) \(\mu=2\). (d)–(f) Zero temperature phase diagrams as a function of \(J\) and \(\mu\) for three different interaction strength \(V_1\) values: (d) \(V_1=1\), (e) \(V_1=2\), and (f) \(V_1=3\). In each panel, the annotations indicate the following: \(s\) denotes the \(s\)-wave phase; \((s, es)\) represents a mixed phase of \(s\)-wave and \(es\)-wave pairings; \(d+id\) refers to the chiral \(d\)-wave phase (specifically with \(d+id\) pairing); \(f\) signifies the \(f\)-wave phase; \(p+ip\) indicates the chiral \(p\)-wave phase (with \(p+ip\) pairing); and \((f, p+ip)\) denotes a mixed phase of \(f\)-wave and \(p+ip\)-wave pairings. These phase diagrams reveal the emergence of chiral \(d\)-wave—or the chiral \(p\)-wave phase (\(p+ip\)-wave pairing). The white region corresponds to the normal phase (\(N\)), where all the pairings are vanishing. 
    In panels (c)–(f), the red boxes around \(J \approx 0.6\) and \(\mu \approx 2.0\) indicate the relevant parameter space for the candidate material \(\mathrm{CrSb}\) \cite{Reimers2024}. 
    }
    \label{fig:phase_diagram}
\end{figure*}

We determine the energetically favored superconducting states over a range of parameters—\(\mu\), \(J\), and \(V_1\)—by numerically solving the self-consistent gap equations at zero temperature [Eqs.~\eqref{eq:self_cons_eq_singlet} and \eqref{eq:self_cons_eq_triplet}]. All possible superconducting states, including both spin-singlet and spin-triplet configurations, as well as their pure and mixed variants, are considered in the analysis. The condensation energy for each state is calculated and compared to identify the ground state. We use \(U=3\) and \(V_2=0\) throughout the main text, while the case of \(V_2 > 0\) is explored in the Appendix. Methodologies are detailed in Appendices \ref{app:BdG_analysis_singlet} and \ref{app:BdG_analysis_triplet}.

Figure~\ref{fig:phase_diagram}(a) displays the phase diagram as a function of \(J\) and \(V_1\) at \(\mu=-2\) and \(V_2=0\). For small \(J\), the system energetically favors spin singlet states, since electrons with opposite spins and momenta—forming singlet pairs—have nearly equal energies (\(E_{\alpha +}(\bm{k}) \approx E_{\alpha -}(-\bm{k})\)). The specific pairing wave symmetry depends on the values of \(V_1\): a pure \(s\)-wave phase appears in \(V_1 < 0.8\), while a mixture of \(s\)-wave and \(es\)-wave components—referred to as the \((s, es)\)-wave phase—arises for \(V_1 \geq 0.8 \). On the other hand, at large \(J\), spin triplet states emerge as the ground state, supported by the energy degeneracy (\(E_{\alpha +}(\bm{k})=E_{\alpha +}(-\bm{k})\)) of electrons in triplet pairs—electrons with the same spin and opposite momenta—which is maintained regardless of \(J\). The detailed pairing wave symmetry depends on the values of both \(V_1\) and \(J\): a pure \(f\)-wave phase predominates in the intermediate \(J\) regime, while a pure \(p+ip\)-wave phase, referred to as a chiral \(p\)-wave phase, emerges for sufficiently large \(J\). Additionally, in the region between these phases, as well as when \(V_1\) and \(J\) are both large, a mixed phase of the two wave characteristics, denoted as the \((f, p+ip)\)-phase, occurs. A normal state appears at small \(V_1\) and large \(J\) (approximately within the regime defined by \(V_1 \lesssim 0.6\) and \(J \gtrsim 1.2)\).

Varying \(\mu\) produces qualitatively distinct phase diagrams in the \(J\)–\(V_1\) plane. In particular, at \(\mu=0\) a chiral \(d\)-wave phase emerges at \(V_1\geq3\) and \(J \leq 0.4\) , while the 
\((s, es)\)-wave region in the spin-singlet sector shrinks in favor of an expanded pure \(s\)-wave phase.
[Fig.~\ref{fig:phase_diagram}(b)]. Furthermore, the \((f, p+ip)\)-wave phase emerges as the dominant phase in the spin-triplet domain due to its expansion alongside the contraction of the \(f\)-wave and chiral \(p\)-wave phases. At \(\mu=2\), both the chiral \(d\)-wave and \(f\)-wave phases vanish from the phase diagram, whereas the \((f, p+ip)\)-wave phase is restricted to smaller areas with high \(V_1\) and \(J\) values [Fig.~\ref{fig:phase_diagram}(c)]. Meanwhile, the chiral \(p\)-wave phase expands across a broader \(J\)–\(V_1\) region, becoming the dominant triplet phase. We have confirmed that these three phase diagrams remain qualitatively valid across a broader range of \(\mu\) values: Fig. \ref{fig:phase_diagram}(a) applies for \(-2.3 \leq \mu \leq -1.3\), Fig. \ref{fig:phase_diagram}(b) for \(-0.5 \leq \mu \leq 0.3\), and Fig. \ref{fig:phase_diagram}(c) for \(1.7 \leq \mu \leq 2.3\), with major changes primarily occurring as variations in the areas of each phase. The remaining ranges of \(\mu\), specifically \(-1.3 < \mu < -0.5\) and \(0.3 < \mu < 1.7\), exhibit two different patterns that retain overall similarities to those in Figs. \ref{fig:phase_diagram}(a)–(c), with some notable discrepancies. Notably, the chiral \(d\)-wave phase appears throughout the range of \(-1.1 \leq \mu \leq 0.3\) in the high \(V_1 \gtrsim 3\) and low \(J \lesssim 0.4\) regime. Further details can be found in Appendix \ref{app:phase_diagrams}. In total, the three patterns established here, along with the two additional patterns discussed in the appendix, constitute five distinctive patterns of superconducting phase distribution in our system within the \(J\)–\(V_1\) space when \(V_2\) is disregarded.

To better understand the influence of \(\mu\), we examine phase diagrams as a function of \(J\) and \(\mu\) while fixing \(V_1\) at several values. Figure~\ref{fig:phase_diagram}(d) displays the \(J\)–\(\mu\) phase diagram at \(V_1 = 1\). For \(J \lesssim 0.6\) the singlet phases are established, with the pure \(s\)-wave phase occupying the range \( -1 < \mu < 1.2\), whereas the mixed \((s, es)\)-wave phase appears for \( \mu < -1 \) and \( \mu > 1.2 \). At the smallest \(\mu=-2.5\), the \(f\)-wave phase occupies the entire \(J\) range of the spin triplet phase (\(0.5\lesssim J\leq2\)), with the chiral \(p\)-wave phase being absent in this range. As \(\mu\) increases, the range of the chiral \(p\)-wave phase continuously expands, allowing it to occupy the entire \(J\) range at \(\mu \approx -1\), accompanied by the complete disappearance of the \(f\)-wave phase. For \(\mu \gtrsim -1\), the mixed \((f, p+ip)\)-wave phase begins to emerge at intermediate \( J \approx 0.8 \) and its \(J\) range expands as \(\mu\) increases. At around \(\mu = 0.0\) and up to \(\mu = 0.5\), the pure \(f\)-wave phase re-emerges in the whole triplet regime (\(J \geq 0.7\)), whereas for larger \(\mu\) the chiral \(p\)-wave phase reappear at first for intermediate \(J\) values and its area increases with further increase of \(\mu\), until for \(\mu \geq 1.5\) only the chiral \(p\)-wave phase remains for \( J \geq 0.6\). At \(V_1 = 2\), a similar alternating pattern of \(f\)-wave, chiral \(p\)-wave and the mixed \((f, p+ip)\)-wave phases is observed [Fig. \ref{fig:phase_diagram}(e)]. At \(V_1 = 3\), this alternating pattern remains, but the expansion of the \((f, p+ip)\)-wave phase significantly reduces the areas of the \(f\)-wave and chiral \(p\)-wave phases in the intermediate \(\mu\) range \((-1.5 \lesssim \mu \lesssim 1.5)\) [Fig. \ref{fig:phase_diagram}(f)]. Notably, the chiral \(d\)-wave phase arises near \(\mu \approx 0\) when \(J\) is small and \(V_1\) is relatively large. Additionally, the spin singlet phase disappears across the entire \(J\) range in certain \(\mu\) ranges. We highlight that for all \(V_1\) values, the chiral \(p\)-wave phase consistently emerges across various \(\mu\) ranges, provided that \(J\) is sufficiently strong \((J \gtrsim 0.5)\) [Figs. \ref{fig:phase_diagram}(d)–(f)].

Finally, we discuss the influence of $V_2$ on the superconducting phase diagrams. To this end, we have calculated the $J$--$V_1$ phase diagrams for several finite values of $V_2$ while keeping the other parameters consistent with those in Figs.~\ref{fig:phase_diagram}(a)--(c). Our findings demonstrate that the primary impact of $V_2$ is the emergence of a new spin-singlet phase, specifically the $(s,s_z)$-wave state, which consists of a mixture of $s$-wave and $s_z$-wave pairing amplitudes. For $V_2 > 2$, the $(s,s_z)$-wave phase supplants the previously dominant $s$-wave and $(s,es)$-wave phases. Notably, the region occupied by the chiral $d$-wave phase remains unaffected by $V_2$. Furthermore, the spin-triplet phases, including the chiral $p$-wave state, remain largely stable; they are only marginally suppressed in the regime of large $V_2 > 2$ and relatively small $J$. This overall robustness arises because the $p_z$-wave pairing amplitude remains consistently zero across all explored parameter regimes. Consequently, we conclude that both the chiral $d$-wave and $p$-wave phases are significantly robust against the influence of $V_2$. A detailed analysis and the corresponding phase diagrams are provided in Appendix~\ref{app:phase_diagrams}.

To explore a plausible physical scenario for the candidate material CrSb, we focus on the parameter space within the ranges $0.35 \leq J \leq 0.85$ and $1.6 \leq \mu \leq 2.4$, noting that $J \approx 0.6$ and $\mu \approx 2$ have been shown to accurately describe its electronic band structure \cite{Reimers2024}. For $V_1 > 1$, we find that the chiral $p$-wave phase emerges for larger values of $J$ ($J \gtrsim 0.6$), while the $(s, es)$-wave phase is stabilized for $J \lesssim 0.6$ [Fig. \ref{fig:phase_diagram}(c)]. Notably, the competition between these two phases remains largely independent of $\mu$ and $V_1$ within the explored range [Figs. \ref{fig:phase_diagram}(d)–(f)]. Consequently, our results suggest that CrSb can host chiral $p$-wave pairing for these moderate values of $J$, $\mu$, and $V_1$.

In summary, our mean-field analysis, presented in Fig. \ref{fig:phase_diagram}, demonstrates that the chiral \(p\)-wave phase is stabilized by sufficiently strong coupling \(J \gtrsim 0.6\) across a wide range of \((V_1, \mu)\) values. Additionally, chiral \(p\)-wave pairing can also manifest in a mixed form within the \((f, p + ip)\)-wave phase in other \((V_1, \mu)\) regions. Furthermore, the chiral \(d\)-wave phase can be stabilized within \(-1.1\leq\mu\leq0.3\) by tuning \(J\) to lower values and \(V_1\) to relatively higher values. Notably, all these phases remain stable for \(V_2 < 3\).

\subsection{Distribution of pairing amplitude components} \label{sec:pairing_amplitude}

\begin{figure*}[t]
    \centering
    \includegraphics[width=0.99\linewidth]{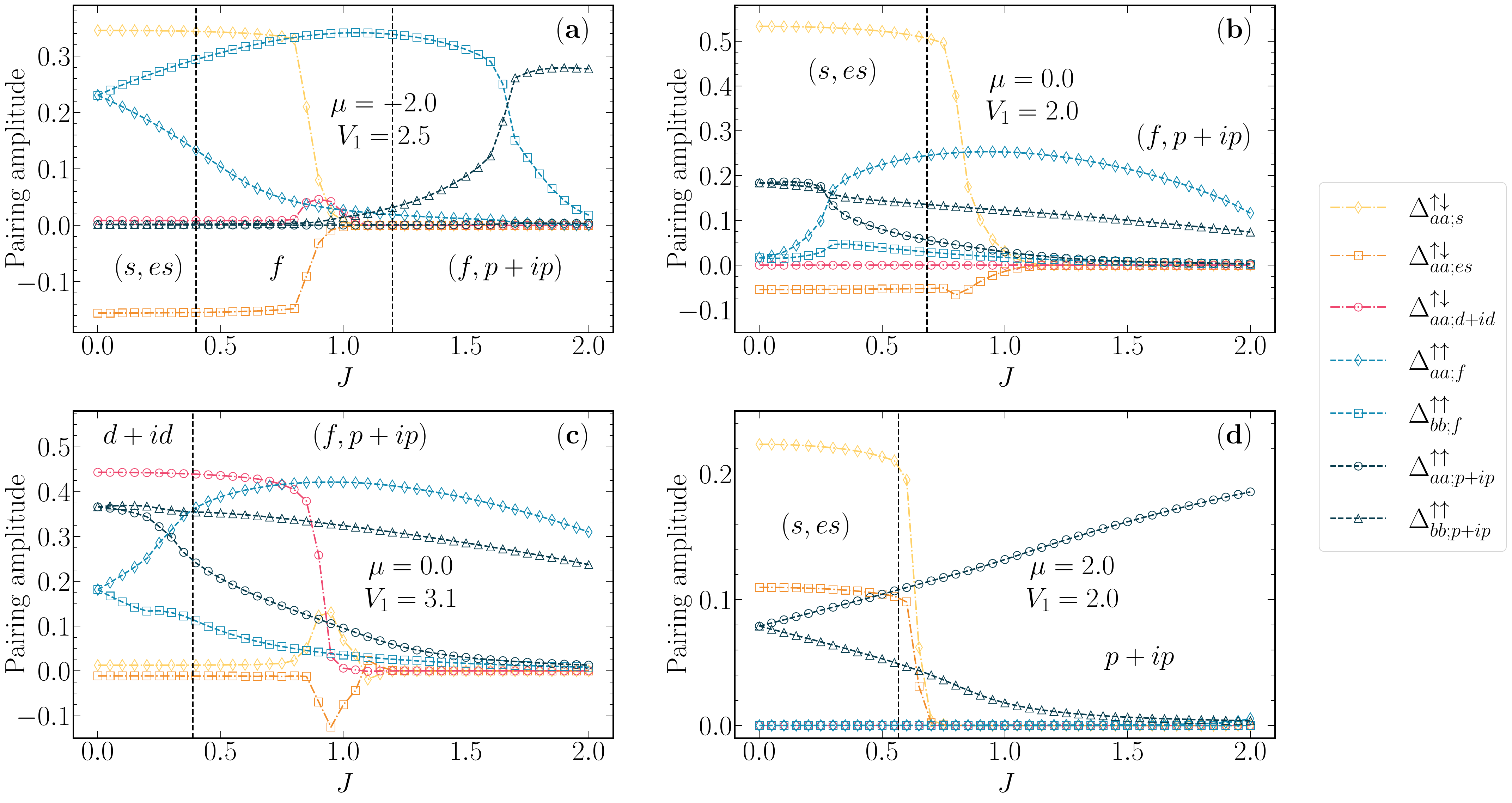}
    \caption{
    \textbf{Distribution of pairing amplitude components.} Light colored markers represent three pairing amplitude components for a spin singlet state as a function of \(J\): \(s\)-wave \((\Delta_{aa;s}^{\uparrow\downarrow})\) (yellow), \(es\)-wave (\(\Delta_{aa;es}^{\uparrow\downarrow}\)) (orange), chiral \(d\)-wave (\(\Delta_{aa;d+id}^{\uparrow\downarrow}\)) (red) pairing amplitudes on the A sublattice. Blue markers represent four pairing amplitude components for a spin triplet state: \(f\)-wave (\(\Delta_{aa;f}^{\uparrow\uparrow}\) and \(\Delta_{bb;f}^{\uparrow\uparrow}\)) and chiral \(p\)-wave (\(\Delta_{aa;p+ip}^{\uparrow\uparrow}\) and \(\Delta_{bb;p+ip}^{\uparrow\uparrow}\)) pairing amplitudes on the A and B sublattices. At each \(J\) point, the pairing amplitude components are derived from their respective spin singlet and spin triplet energy-minimized configurations. The actual ground state is determined by selecting the configuration with the lower condensation energy. Each panel annotates the stabilized ground state phases within their respective regions separated by dashed lines. Parameter values for \(\mu\) and \(V_1\) are annotated in each panel. 
    } 
    \label{fig:pairing_amplitudes}
\end{figure*}

Figure \ref{fig:pairing_amplitudes} illustrates the distribution of three pairing amplitude components for a spin singlet state: \(s\)-wave \((\Delta_{aa;s}^{\uparrow\downarrow})\), \(es\)-wave (\(\Delta_{aa;es}^{\uparrow\downarrow}\)), chiral \(d\)-wave (\(\Delta_{aa;d+id}^{\uparrow\downarrow}\)) pairing amplitudes on the A sublattice as well as four pairing amplitude components for a spin triplet state: \(f\)-wave (\(\Delta_{aa;f}^{\uparrow\uparrow}\) and \(\Delta_{bb;f}^{\uparrow\uparrow}\)) and chiral \(p\)-wave (\(\Delta_{aa;p+ip}^{\uparrow\uparrow}\) and \(\Delta_{bb;p+ip}^{\uparrow\uparrow}\)) pairing amplitudes on the A and B sublattices. The remaining components are related to these components, as detailed in the fourth paragraph of Sec. \ref{secIIB}. The \((s, es)\)-wave phase at \((\mu, V_1) = (-2, 2.5)\) displays largely constant values for \(\Delta_{aa;s}^{\uparrow\downarrow}\) and \(\Delta_{aa;es}^{\uparrow\downarrow}\) regardless of \(J\) [Fig. \ref{fig:pairing_amplitudes}(a)]. Additionally, the inequality \(\Delta_{aa;s}^{\uparrow\downarrow} > \Delta_{aa;es}^{\uparrow\downarrow}\) is maintained, indicating that \(s\)-wave pairing is the predominant type in this phase. The same pattern is observed in the \((s, es)\)-wave phase at both \((\mu, V_1) = (0, 2)\) and \((\mu, V_1) = (2, 2)\) [Figs. \ref{fig:pairing_amplitudes}(b) and (d)]. At \((\mu, V_1) = (-2, 2.5)\), the chiral \(d\)-wave phase appears in place of the \((s, es)\)-wave phase at the small \(J\) regime [Fig. \ref{fig:pairing_amplitudes}(c)]. This phase features a substantial value for \(\Delta_{aa;d+id}^{\uparrow\downarrow}\), which remains largely constant regardless of \(J\), while \(\Delta_{aa;s}^{\uparrow\downarrow}\) and \(\Delta_{aa;es}^{\uparrow\downarrow}\) exhibit negligible values, both of the order of 0.01.

At \((\mu, V_1) = (-2, 2.5)\), two spin triplet phases are identified: the \(f\)-wave phase for \(0.4 < J \leq 0.9\) phase and the \((f, p+ip)\)-wave phase for \(J > 0.9\) phase [Fig. \ref{fig:pairing_amplitudes}(a)]. 
The \(f\)-wave phase shows a decrease in \(\Delta_{aa;f}^{\uparrow\uparrow}\) with increasing \(J\) and an increase in \(\Delta_{bb;f}^{\uparrow\uparrow}\), making \(\Delta_{bb;f}^{\uparrow\uparrow}\) the dominant component. In the \((f, p + ip)\)-wave phase, nonzero values of \(\Delta_{bb;p+ip}^{\uparrow\uparrow}\) complement the existing components \(\Delta_{aa;f}^{\uparrow\uparrow}\) and \(\Delta_{bb;f}^{\uparrow\uparrow}\). Notably, \(\Delta_{aa;f}^{\uparrow\uparrow}\) keeps decreasing and \(\Delta_{bb;f}^{\uparrow\uparrow}\) begins to decline at \(J=1.1\) whereas \(\Delta_{bb;p+ip}^{\uparrow\uparrow}\) keeps increasing. These trends ultimately result in \(\Delta_{bb;p+ip}^{\uparrow\uparrow}\) becoming the dominant type in the large \(J > 2\) regime. At both \((\mu, V_1) = (0, 2)\) and \((\mu, V_1) = (0, 3.1)\), the \((f, p+ip)\)-wave phase exhibits an overall declining trend across its pairing amplitude components: \(\Delta_{aa;f}^{\uparrow\uparrow}\), \(\Delta_{bb;f}^{\uparrow\uparrow}\), \( \Delta_{aa;p+ip}^{\uparrow\uparrow}\), and \( \Delta_{bb;p+ip}^{\uparrow\uparrow}\), although a slight uptick is observed for \(\Delta_{aa;f}^{\uparrow\uparrow}\) at lower \(J\) values [Figs. \ref{fig:pairing_amplitudes}(b) and (c)]. Furthermore, \(\Delta_{aa;f}^{\uparrow\uparrow}\) remains the largest component throughout this phase, making this \(f\)-wave pairing the predominant type. Finally, the chiral \(p\)-wave phase at \((\mu, V_1) = (2, 2)\) exhibits increasing and decreasing trends for \(\Delta_{aa;p+ip}^{\uparrow\uparrow}\) and \(\Delta_{bb;p+ip}^{\uparrow\uparrow}\), respectively, making \(\Delta_{aa;p+ip}^{\uparrow\uparrow}\) the dominant component [Fig. \ref{fig:pairing_amplitudes}(d)].

We emphasize that the pairing amplitudes of the \((s, es)\)-wave and chiral \(d\)-wave states collapse within specific intermediate \(J\) regimes: \(0.8 \leq J \leq 1.0\) for panel (a), \(0.7 \leq J \leq 1.1\) for panel (b), \(0.8 \leq J \leq 1.0\) for panel (c), and \(0.6 \leq J \leq 0.7\) for panel (d) in Fig. \ref{fig:pairing_amplitudes}. Beyond the upper limits of these ranges, the pairing amplitudes vanish entirely, indicating an absence of stable spin singlet solutions in the large \(J\) regime. In contrast, the pairing amplitudes of the \(f\)-wave, chiral \(p\)-wave, and \((f, p+ip)\)-wave states remain robust beyond the upper limits. Consequently, these spin triplet states manifest as the stable superconducting phases in the large \(J\) regime.

\subsection{Suppression of spin singlet phases}

\begin{figure*}[t!]
    \centering       
    \includegraphics[width=\linewidth]{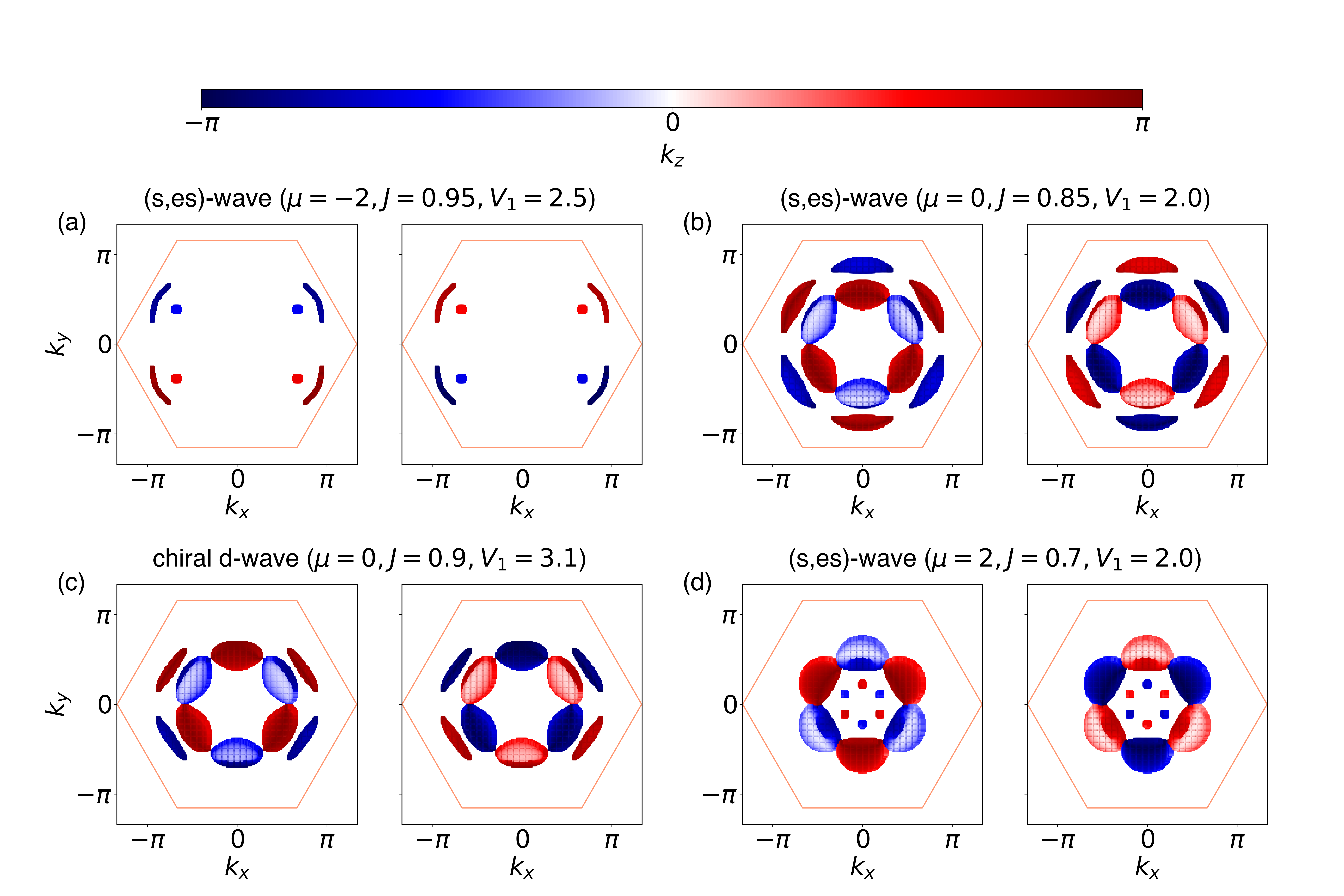}
    \caption{\textbf{Formation of Bogoliubov Fermi surfaces (BFSs).} In each panel, the colored areas represent the \(\bm{k}\)-space volume where BFSs are present. The \(\bm{k}\)-points are projected onto the \(k_x\)–\(k_y\) plane, with the \(k_z\) value indicated by the color scale. Each panel corresponds to an \((s, es)\)-wave or chiral \(d\)-wave state, with specific parameter values annotated. The left and right subpanels in each panel represent the third and fourth lowest energy bands, respectively. The \(x\)-axis indicates the \(k_x\) values, while the \(y\)-axis represents the \(k_y\) values in all panels.
    }
    \label{fig:BFS}
\end{figure*}

The formation of BFSs in the spin singlet states plays a crucial role in the collapse of pairing amplitudes of spin singlet states in the intermediate \(J\) regimes, as discussed in the third paragraph of Sec. \ref{sec:pairing_amplitude}. To illustrate this, we examine the quasiparticle energy bands of spin singlet states—the four eigenenergies of \(\hat{H}_\textrm{BdG}^\textrm{singlet}\), as presented in Eq. \eqref{eq:H_BdG_singlet}, as a function of \(\bm{k}\). We consider two different \(J\) regimes: the small \(J\) regime, where the pairing amplitudes of spin singlet states remain substantial, and the intermediate \(J\) range, where these amplitudes collapse. In the small \(J\) regime, all energy bands exhibit positive values throughout the entire Brillouin zone (BZ) and show no signs of BFSs. However, in the intermediate \(J\) range, the two lower bands display negative energy levels, leading to the formation of BFSs in specific \(\bm{k}\) regions, while the two upper bands exhibit no signatures of BFSs. Specifically, in the \((s,es)\)-wave state, the BFSs of the two lower bands occur in the \(k_x\)–\(k_y\) regions where the influence of the altermagnetic fields is strongest, corresponding to the maxima of \(t_z(\bm{k})\) [Figs. \ref{fig:BFS}(a), (b), and (d)]. Additionally, these BFSs are located in the \(k_z>0\) region where the sign of \(t_z(\bm{k})\) is positive, and vice versa. The chiral \(d\)-wave state also displays BFSs in the strongest field regions; however, their placement in the \(k_z\) axis is inverted [Fig. \ref{fig:BFS}(c)], attributed to the opposite sign of the form factor of this state, as displayed in Eq. \eqref{eq:g_k_singlet}. In contrast, our numerical analysis has found that all of the quasiparticle energy bands of spin triplet states, described by the four eigenenergies of \(\hat{H}_\textrm{BdG}^\textrm{triplet}\) in Eq. \eqref{eq:H_BdG_triplet}, uniformly show positive values with no indication of BFSs throughout the full range of \(J\).

The formation of BFSs in the spin singlet states leads to the vanishing of the spin-singlet pair correlation functions, \(\langle c_{-\bm{k}s\uparrow} c_{\bm{k}s\downarrow} \rangle\) for \(s = a, b\), in the regions of \(\bm{k}\) space where BFSs occur [Figs. \ref{fig:pair_correlation_function}(a)–(d)]. This results in the exclusion of these \(\bm{k}\) space areas from momentum integration in the self-consistent gap equations, thereby suppressing the pairing amplitudes of the spin singlet states. In contrast, the spin-triplet pair correlation functions, \(\langle c_{-\bm{k}s\sigma} c_{\bm{k}s\sigma} \rangle\) for \(s = a, b\) and \(\sigma = \uparrow, \downarrow\), remain robust across a significant portion of the \(\bm{k}\) space due to the absence of BFSs in the spin triplet states. However, the vanishing of spin-triplet form factors along specific high-symmetry lines still restricts the available phase space volume for momentum integration in the self-consistent gap equations [Figs. \ref{fig2:harmonics}(e)–(h)]. These two effects interact to determine which pairing type—spin singlet or spin triplet—dominates. When \(J\) is small, the restrictions imposed by the triplet form factors have a greater impact than the diminishing spin-singlet pair correlations, resulting in spin singlet pairing as the dominant type, as shown in Fig. \ref{fig:pairing_amplitudes}. Conversely, for large \(J\), the impact of the diminishing spin-singlet correlations prevails, stabilizing spin triplet states as the superconducting phases in the high \(J\) regime. Furthermore, it is likely that the interplay between these two effects determines the critical value of \(J\) for the phase transition between the spin singlet and triplet phases.

To better understand the formation of BFSs in the spin singlet states, we derive the energy dispersions of BdG quasiparticles in these states as follows:
\begin{equation}
\label{eq:quasiparticle_energy_singlet}
\Xi_{\alpha\beta}^\textrm{singlet}(\bm{k}) = \Bigg[\epsilon_0(\bm{k})^{2}+t_{x}(\bm{k})^{2}+t_{z}(\bm{k})^{2}+J^{2}+|\Delta_{\uparrow\downarrow}(\bm{k})|^{2} + \alpha \frac{2\epsilon_0(\bm{k})t_{z}(\bm{k})J}{\sqrt{S(\bm{k})}}-S(\bm{k})\Bigg]^{\frac{1}{2}} + \beta \sqrt{S(\bm{k})}.
\end{equation}
In this expression, \(\alpha = \pm\) and \(\beta = \pm\) denote the four bands; \(\Delta_{\uparrow\downarrow}(\bm{k})=\Delta_{aa}^{\uparrow\downarrow}(\bm{k}) = \Delta_{bb}^{\uparrow\downarrow}(\bm{k})\) stands for the pairing gap function; \(S(\bm{k})\) is a function composed of \(\epsilon_0(\bm{k})\), \(|\Delta_{\uparrow\downarrow}(\bm{k})|\), \(t_x(\bm{k})\), \(t_z(\bm{k})\), and \(J\). The derivation of \(\Xi_{\alpha\beta}^\textrm{singlet}(\bm{k})\) and the detailed expression of \(S(\bm{k})\) can be found in Appendix \ref{app:BdG_analysis_singlet}. Notably, the two lower bands of \(\Xi_{\pm-}^\textrm{singlet}(\bm{k})\) exhibit a negative sign for the second term, \(\beta \sqrt{S(\bm{k})}\), and a positive sign for the first term in the bracket. This indicates that when the second term exceeds the first for certain \(\bm{k}\) values, the lower bands can become negative and lead to the formation of BFSs in the corresponding \(\bm{k}\) regions. In contrast, the two upper bands \(\Xi_{\pm+}^\textrm{singlet}(\bm{k})\) display positive values for all \(\bm{k}\) values, as both the first and second terms are positive. Our findings have confirmed that increasing \(J\) enhances the second term and leads to the formation of BFSs in the lower bands in the high \(J\) regime.

The physical origin of the term \(\beta \sqrt{S(\bm{k})}\) is attributed to the altermangetic spin splitting effect caused by \(J\). To illustrate this, we consider the simplified energy dispersion for the case \(t_3=0\), where \(t_x(\bm{k})=0\) and the term \(\beta\sqrt{S(\bm{k})}\) reduces to \(\beta J\): \(\tilde{\Xi}_{\alpha\beta}^\textrm{singlet}(\bm{k}) = \big[ (\epsilon_0(\bm{k}) + \alpha t_z(\bm{k}))^2 + |\Delta_{\uparrow\downarrow}(\bm{k})|^2 \big]^{\frac{1}{2}} + \beta J \). In this expression, a large value of \(J\) enables the lower bands, \(\tilde{\Xi}_{\pm-}^\textrm{singlet}(\bm{k})\), to accommodate BFSs within the \(\bm{k}\) regions, where \(\big[ (\epsilon_0(\bm{k}) + \alpha t_z(\bm{k}))^2 + |\Delta_{\uparrow\downarrow}(\bm{k})|^2 \big]^{\frac{1}{2}} < J\). Furthermore, it is evident that the term \(\beta J\) represents the altermagnetic spin splitting effect, which arises from the opposite energy shifts of two electrons with opposing spins. This is illustrated by \(\frac{\beta}{2}\big(E_{\alpha+}(\bm{k})-E_{\alpha-}(-\bm{k})\big) = \beta J \), where \(E_{\alpha\beta}(\bm{k}) = \epsilon_0(\bm{k}) + \alpha (t_{z}(\bm{k}) + \beta J)\). Similarly, \(\beta \sqrt{S(\bm{k})}\) reflects a comparable splitting effect in the case of \(t_3=0.6\), although the mixing of hopping energy between the two different sublattices due to \(t_x(\bm{k})\) complicates its expression. Therefore, the altermagnetic spin splitting effect is a key mechanism for the occurrence of BFSs in the spin singlet states.

The absence of BFSs in spin triplet states can be understood from the energy dispersions of BdG quasiparticles in these states, which are given as follows (derivation can be found in Appendix \ref{app:BdG_analysis_triplet}):
\begin{equation}
\begin{aligned}
\label{eq:quasiparticle_energy_triplet}
\Xi_{\alpha\sigma}^\textrm{triplet}(\bm{k}) = & \; \bigg[\epsilon_0(\bm{k})^2+t_x(\bm{k})^2+(t_z(\bm{k})+s(\sigma)  J)^2+|\Delta_{+}^\sigma(\bm{k})|^2+|\Delta_{-}^\sigma(\bm{k})|^2 \\
&+2\alpha \sqrt{\big(\epsilon_0(\bm{k})(t_z(\bm{k})+s(\sigma)  J)+|\Delta_{+}^\sigma(\bm{k})||\Delta_{-}^\sigma(\bm{k})|\big)^2+t_x(\bm{k})^2\big(\epsilon_0(\bm{k})^2+|\Delta_{-}^\sigma(\bm{k})|^2\big)} ~ \bigg]^{\frac{1}{2}}.
\end{aligned}
\end{equation}
In this expression, \(\alpha = \pm\) denote the upper and lower bands; \(\sigma = \; \uparrow, \downarrow \) signify up and down spins; the pairing gap functions are given by \(\Delta_{\pm}^\sigma(\bm{k}) = \frac{1}{2}[\Delta_{aa}^{\sigma\sigma}(\bm{k}) \pm \Delta_{bb}^{\sigma\sigma}(\bm{k})]\). Evidently, all four bands of \(\Xi_{\alpha\sigma}^\textrm{triplet}(\bm{k})\) uniformly show positive values with no indication of BFSs throughout the full range of \(J\). This behavior is consistent with our earlier observation based on the solution of the self-consistent gap equations, as discussed in the first paragraph of this section. The reason for this behavior is that unlike the spin singlet pairing, the overall energy in the quasiparticle energy bands is identically zero, because two electrons having the same spin in a triplet pair remain at the same energy level despite the altermagnetic field, as shown by \(E_{\alpha+}(\bm{k})-E_{\alpha+}(-\bm{k}) = 0\). As a result, the quasiparticle energy bands exhibit either hard-gapped spectra or pseudo-gapped spectra with nodal points or lines, showing no indication of BFSs in the spin triplet states.

In summary, the spin singlet states undergo the formation of BFSs due to the altermagnetic spin splitting effect, which results in a loss of energetic stability in the high \(J\) regime. In contrast, the spin triplet states are unaffected by the altermagnetic spin splitting effect, allowing them to remain a stable superconducting phase.

\subsection{Experimental signatures} \label{sec:exp_sign}

\begin{figure}
\includegraphics[width=\linewidth]{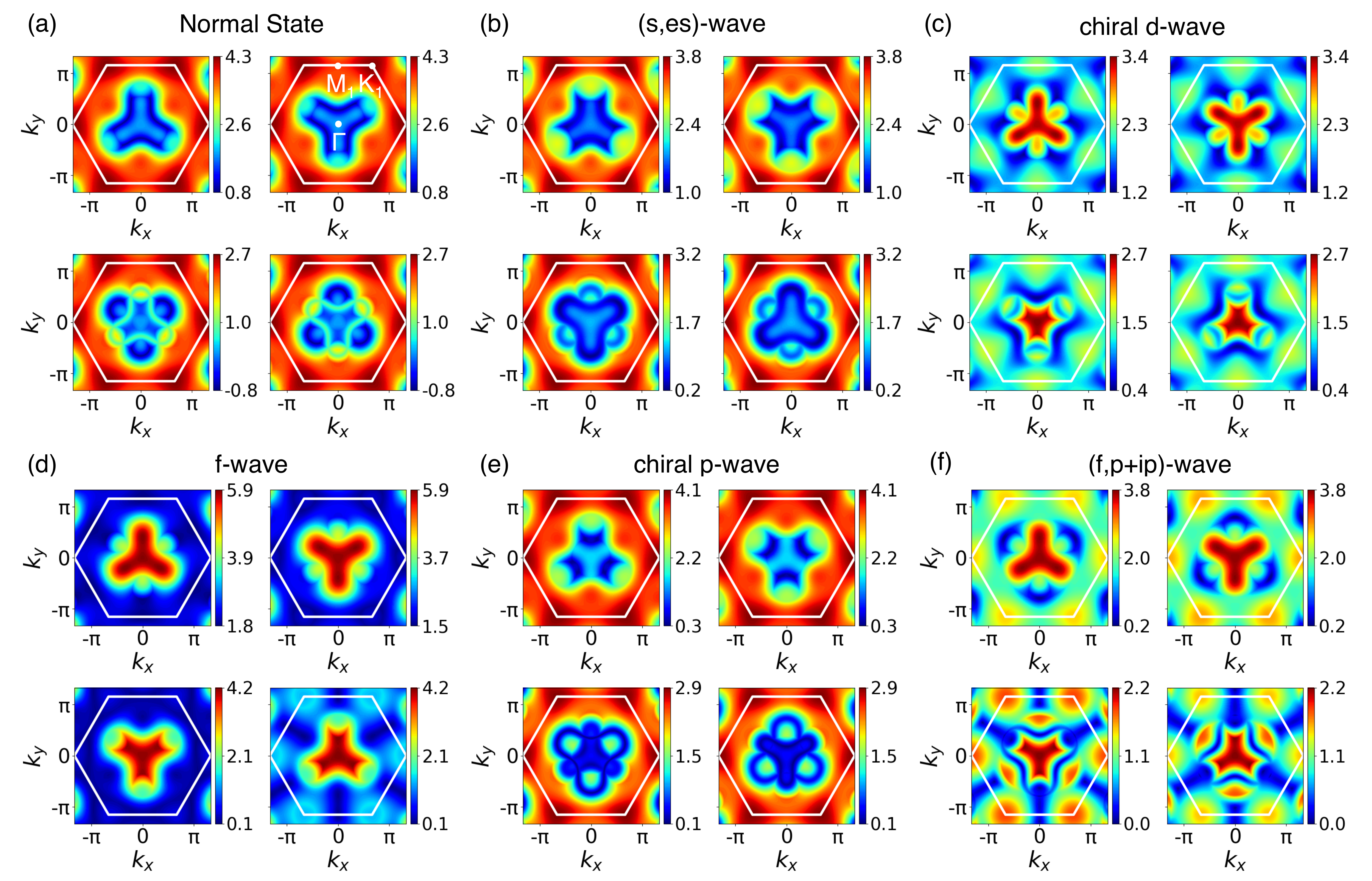}
\centering
\caption{\textbf{Energy dispersions of BdG quasiparticles.}
(a) Normal-state electron energy dispersions for $(\mu, J) = (2, 0.8)$. (b)--(f) Energy dispersions of Bogoliubov--de Gennes (BdG) quasiparticles for the following pairing states: (b) $(s, es)$-wave, (c) chiral $d$-wave, (d) $f$-wave, (e) chiral $p$-wave, and (f) mixed $(f, p+ip)$-wave. The parameter sets $(\mu, V_1, J)$ for each panel are: (b) $(2, 0.3, 2)$, (c) $(0, 0.3, 3.1)$, (d) $(-2, 0.8, 2.5)$, (e) $(2, 0.8, 2)$, and (f) $(0, 0.8, 3.1)$. In all panels, the four subpanels represent the spin and band sectors: upper-left (spin-up upper band), upper-right (spin-down upper band), lower-left (spin-up lower band), and lower-right (spin-down lower band). All energy dispersions are projected onto the $k_x$--$k_y$ plane at $k_z = \pi/2$.}
\label{fig:energy_dispersion}
\end{figure}

We investigate the BdG quasiparticle energy dispersions as experimental signatures to identify the specific symmetry-breaking patterns induced by the proposed pairing states. In the normal state, the electron energy dispersions exhibit $\mathcal{C}_{3z}$ symmetry, and the spin-up and spin-down sectors are related by reflection symmetry across the $k_x$-axis, as shown in Fig.~\ref{fig:energy_dispersion}(a). The energy dispersions for the $(s,es)$-wave state preserve both of these symmetries, as expected from its fully symmetric form factor [Fig.~\ref{fig:energy_dispersion}(b)]. Similarly, the dispersions for the chiral $d$-wave and chiral $p$-wave states also preserve both symmetries [Figs.~\ref{fig:energy_dispersion}(c) and (e)]. This preservation is non-trivial because the real and imaginary components of their respective form factors individually break $\mathcal{C}_{3z}$ symmetry while respecting the reflection symmetry [Figs. \ref{fig2:harmonics}(c), (d), (g), and (h)]. Nevertheless, the resulting energy dispersions maintain $\mathcal{C}_{3z}$ symmetry. This stems from the fact that the form factors contribute to the quasiparticle energy expressions [Eqs.~\eqref{eq:quasiparticle_energy_singlet} and \eqref{eq:quasiparticle_energy_triplet}] only via their absolute values, which are themselves $\mathcal{C}_{3z}$-invariant, as illustrated in Figs.~\ref{fig2:harmonics}(b) and (f). The energy dispersions of the $f$-wave state preserve $\mathcal{C}_{3z}$ symmetry but break the reflection symmetry, consistent with the intrinsic symmetry of its form factor [Fig.~\ref{fig:energy_dispersion}(d)]. Furthermore, these dispersions exhibit minima along the high-symmetry lines, such as $M_1$--$\Gamma$--$M_1$ and its symmetry-related counterparts, reflecting the fact that the $f$-wave form factor vanishes along these directions.

The energy dispersions of the $(f, p+ip)$-wave state break $\mathcal{C}_{3z}$ symmetry [Fig.~\ref{fig:energy_dispersion}(f)]. This symmetry breaking is attributed to the interference between the $f$-wave and chiral $p$-wave pairing components. To illustrate this, we examine the magnitude of the pairing gap functions for this state:
\begin{equation}
\begin{aligned}
    |\Delta_{\pm}^\sigma(\bm{k})|^2 = & \; |\Delta_f^\sigma|^2|g_f(\bm{k})|^2 + |\Delta_{p+ip}^\sigma|^2|g_{p+ip}(\bm{k})|^2 \\
    & + \big[\Delta_f^\sigma (\Delta_{p+ip}^\sigma)^* \pm (\Delta_f^\sigma)^* \Delta_{p+ip}^\sigma\big] g_f(\bm{k})\text{Re}[g_{p+ip}(\bm{k})] \\
    & - i \big[\Delta_f^\sigma (\Delta_{p+ip}^\sigma)^* \mp (\Delta_f^\sigma)^* \Delta_{p+ip}^\sigma \big] g_f(\bm{k})\text{Im}[g_{p+ip}(\bm{k})],
\end{aligned}
\end{equation}
where $\Delta_f^\sigma = \frac{1}{2}(\Delta_{aa;f}^{\sigma\sigma} + \Delta_{bb;f}^{\sigma\sigma})$ and $\Delta_{p+ip}^\sigma = \frac{1}{2}(\Delta_{aa;p+ip}^{\sigma\sigma} + \Delta_{bb;p+ip}^{\sigma\sigma})$. We note that for finite values of $\Delta_f^\sigma$ and $\Delta_{p+ip}^\sigma$, the cross-terms involving the real or imaginary parts of the form factors remain non-zero. Since $g_f(\bm{k})$ is $\mathcal{C}_{3z}$-invariant but $\text{Re}[g_{p+ip}(\bm{k})]$ and $\text{Im}[g_{p+ip}(\bm{k})]$ individually break $\mathcal{C}_{3z}$ symmetry, their product results in an overall $\mathcal{C}_{3z}$ symmetry breaking in the pairing gap function, and consequently, in the energy dispersions. We have confirmed that a similar symmetry-breaking effect occurs when a chiral $d$-wave pairing mixes with other spin-singlet components. Although such mixed singlet states do not emerge in our current results, they are theoretically possible. Given that all pure states and mixed states composed of non-chiral components preserve $\mathcal{C}_{3z}$ symmetry, we conclude that the breaking of $\mathcal{C}_{3z}$ symmetry in the quasiparticle energy dispersions can serve as a hallmark signature for the formation of chiral pairing states when they mix with non-chiral pairing components.

\begin{figure}
    \centering
    \includegraphics[width=\linewidth]{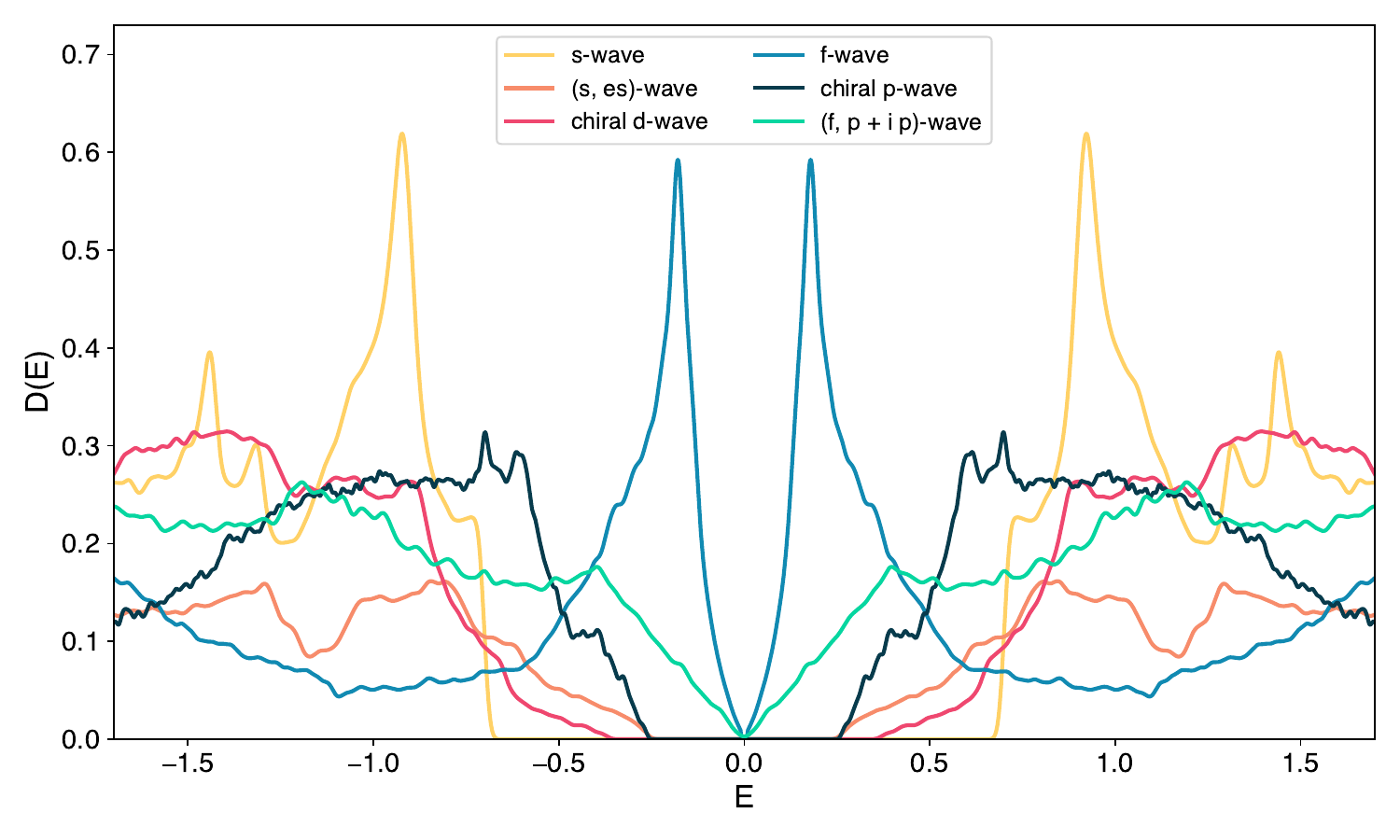}
    \caption{\textbf{Density of states.}
    Density of states, $D(E)$, of the BdG quasiparticle energy dispersions for the six representative superconducting states. The parameter sets for each state correspond to those used in Fig.~\ref{fig:energy_dispersion}, with the addition of the $s$-wave state at $(\mu, J, V_1) = (0, 0.3, 2)$.
    }
    \label{fig:DOS}
\end{figure}

Figure~\ref{fig:DOS} displays the quasiparticle density of states (DOS) for various pairing symmetries. The $(s,es)$-wave, chiral $d$-wave, and chiral $p$-wave states exhibit hard gap structures characterized by a U-shaped profile, reflecting the fact that their pairing amplitudes remain robust across the entire Fermi surface. Notably, although the pairing amplitudes of the chiral $d$-wave and chiral $p$-wave symmetries vanish at the $\Gamma$ point, the gap remains fully open because the Fermi surface does not enclose the $\Gamma$ point at the considered electron filling. Conversely, the $f$-wave and the mixed $(f,p+ip)$-wave states display a V-shaped gap near zero energy. In the case of the $f$-wave state, this occurs because the pairing amplitude vanishes along high-symmetry lines, such as $M_1$--$\Gamma$--$M_1$ and its symmetry-related counterparts. For the mixed $(f,p+ip)$-wave state, the Fermi surface encloses the $\Gamma$ point; since the chiral $p$-wave pairing amplitude vanishes at $\Gamma$, this leads to a suppressed gap and the observed V-shaped feature. Furthermore, we observe a distinction in the coherence peaks: pure states, such as the $s$-wave, chiral $d$-wave, $f$-wave, and chiral $p$-wave symmetries, exhibit sharp coherence peaks at the gap edges. In contrast, the mixed $(s,es)$-wave and $(f, p+ip)$-wave states show a notable absence of such peaks. These distinctive DOS features provide a reliable means of identifying the underlying pairing symmetries.

\section{Conclusion}

Our investigation of the superconducting phase diagram in $g$-wave altermagnetic metals reveals that the chiral $p$-wave phase is stabilized as altermagnetic spin-splitting induces Bogoliubov Fermi surfaces within the spin-singlet sectors, thereby suppressing them. Additionally, the chiral $d$-wave phase is accessible by tuning the chemical potential at intermediate electron densities. These results establish $g$-wave altermagnets as a promising platform for realizing chiral superconductivity. To explore this potential, we propose investigating recently identified $g$-wave altermagnetic candidates, such as CrSb and MnTe. The predicted quasiparticle energy dispersions can be probed via angle-resolved photoemission spectroscopy \cite{Gu2019}. Specifically, the spontaneous breaking of three-fold rotational symmetry serves as a definitive signature of chiral $p$-wave or $d$-wave pairing, including their mixing with non-chiral components. Furthermore, the detailed features of the quasiparticle DOS can be measured using scanning tunneling spectroscopy \cite{RevModPhys.79.353, PhysRevB.101.054505}, which provide additional indicators to distinguish between competing pairing scenarios.

Future research directions include investigating the effects of spin-orbit coupling (SOC), which was omitted in this study for simplicity. In candidate materials such as CrSb and MnTe, SOC is expected to be non-negligible. Its inclusion may induce a mixing of spin-singlet and triplet pairings and lift the degeneracy between helical and chiral phases \cite{PhysRevB.108.184505}. Consequently, incorporating SOC will be vital for a refined understanding of the pairing symmetries in these systems. Another promising avenue involves exploring how altermagnetic fluctuations mediate unconventional pairing. As discussed in Ref.~\citenum{dlpb-gfct}, such fluctuations can generate a rich phase diagram with intra- and inter-unit-cell pairings depending on the proximity to a magnetic phase transition. These future studies will not only deepen our theoretical understanding of $g$-wave altermagnets but also facilitate the application of our findings to real-world quantum materials.

\section*{Acknowledgement}

T.\v{C}. and K.-M.K. were supported by an appointment to the JRG Program at the APCTP through the Science and Technology Promotion Fund and Lottery Fund of the Korean Government. This research was supported by the Korean Local Governments - Gyeongsangbuk-do Province and Pohang City.

\appendix

\section{Derivation of the pairing Hamiltonian}\label{app:H_BdG}

In this section, we present the derivation of spin-singlet and spin-triplet pairing Hamiltonians, which are presented in Eqs.~\eqref{eq:H_BdG_singlet} and \eqref{eq:H_BdG_triplet} of the main text, respectively. The interaction Hamiltonian in Eq.~\eqref{eq:H_int_momentum_space} can be reorganized as:
\begin{equation}
\label{eq:H_p_1}
    \hat{H}_\textrm{int} = - \frac{1}{N} \sum_{\bm{k},\bm{k}',\bm{q}} \sum_{s,s'} \sum_{\sigma,\sigma'} V_{ss'}^{\sigma\sigma'}(\bm{k}+\bm{q}-\bm{k}') c_{\bm{k}+\bm{q}s\sigma}^\dagger c_{-\bm{k}s'\sigma'}^\dagger c_{-\bm{k}'+\bm{q}s'\sigma'} c_{\bm{k}'s\sigma}. 
\end{equation}
For the mean-field approximation, we focus on the $\bm{q}=0$ sector:
\begin{equation}
\label{eq:H_p_2}
    \hat{H}_\textrm{int}^{\bm{q}=0} = - \frac{1}{N} \sum_{\bm{k},\bm{k}'} \sum_{s,s'} \sum_{\sigma,\sigma'} V_{ss'}^{\sigma\sigma'}(\bm{k}-\bm{k}') c_{\bm{k}s\sigma}^\dagger c_{-\bm{k}s'\sigma'}^\dagger c_{-\bm{k}'s'\sigma'} c_{\bm{k}'s\sigma}.
\end{equation}
Using the following mean-field decoupling:
\begin{equation}
\begin{aligned}
\label{eq:mean_field_decoupling}
    & c_{\bm{k}s\sigma}^\dagger c_{-\bm{k}s'\sigma'}^\dagger c_{-\bm{k}'s'\sigma'} c_{\bm{k}'s\sigma} \\
    \approx & \; \langle c_{\bm{k}s\sigma}^\dagger c_{-\bm{k}s'\sigma'}^\dagger \rangle c_{-\bm{k}'s'\sigma'} c_{\bm{k}'s\sigma} + \langle c_{-\bm{k}'s'\sigma'} c_{\bm{k}'s\sigma} \rangle c_{\bm{k}s\sigma}^\dagger c_{-\bm{k}s'\sigma'}^\dagger -\langle c_{\bm{k}s\sigma}^\dagger c_{-\bm{k}s'\sigma'}^\dagger \rangle \langle c_{-\bm{k}'s'\sigma'} c_{\bm{k}'s\sigma} \rangle,
\end{aligned}
\end{equation}
we approximate \(\hat{H}_\textrm{int}^{\bm{q}=0}\) as:
\begin{equation}
\begin{aligned}
\label{eq:H_p_3}
    \hat{H}_\textrm{int}^{\bm{q}=0} \approx & \; - \frac{1}{N} \sum_{\bm{k},\bm{k}'} \sum_{s,s'} \sum_{\sigma,\sigma'} V_{ss'}^{\sigma\sigma'}(\bm{k}-\bm{k}') \Big[\langle c_{\bm{k}s\sigma}^\dagger c_{-\bm{k}s'\sigma'}^\dagger \rangle c_{-\bm{k}'s'\sigma'} c_{\bm{k}'s\sigma} \\
    & \; + \langle c_{-\bm{k}'s'\sigma'} c_{\bm{k}'s\sigma} \rangle c_{\bm{k}s\sigma}^\dagger c_{-\bm{k}s'\sigma'}^\dagger -\langle c_{\bm{k}s\sigma}^\dagger c_{-\bm{k}s'\sigma'}^\dagger \rangle \langle c_{-\bm{k}'s'\sigma'} c_{\bm{k}'s\sigma} \rangle\Big].
\end{aligned}
\end{equation}
The interaction function can be decomposed into: 
\begin{equation}
V_{ss'}^{\sigma\sigma'}(\bm{k}-\bm{k}') = \sum_{\eta} V_{ss';\eta}^{\sigma\sigma'} g_{\eta}(\bm{k}) g_{\eta}^*(\bm{k}'),
\end{equation}
where the summation over \(\eta = \{s, es, d+id, d-id, s_z, f, p+ip, p-ip, p_z\}\) is implied and the interaction coefficients \(V_{ss';\eta}^{\sigma\sigma'}\) are given by 
\begin{equation}
\begin{aligned}
\label{eq:V_coeff}
    V_{ss';s}^{\sigma\sigma'} & = \frac{1}{2}U\delta_{ss'}(1-\delta_{\sigma\sigma'}), \\
    V_{ss';es}^{\sigma\sigma'} & = V_{ss';d\pm id}^{\sigma\sigma'} = V_{ss';f}^{\sigma\sigma'} = V_{ss';p\pm ip}^{\sigma\sigma'} = \frac{1}{3}\delta_{ss'} V_1, \\
    V_{ss';s_z}^{\sigma\sigma'} & = V_{ss';p_z}^{\sigma\sigma'} = \frac{1}{2} V_2(1-\delta_{ss'}).
\end{aligned}
\end{equation} 
Substituting the decomposition of the interaction function into \(\hat{H}_\textrm{int}^{\bm{q}=0}\), we rewrite it in the form:
\begin{equation}
\label{eq:H_p_4}
    \hat{H}_\textrm{int}^{\bm{q}=0} = \sum_{\bm{k}} \sum_{s,s'} \sum_{\sigma, \sigma'} \sum_{\eta} \Big[ \Delta_{ss';\eta}^{\sigma\sigma'}g_{\eta}(\bm{k}) c_{\bm{k}s\sigma}^\dagger c_{-\bm{k}s'\sigma'}^\dagger
    + \left[\Delta_{ss';\eta}^{\sigma\sigma'}g_{\eta}(\bm{k})\right]^* c_{-\bm{k}s'\sigma'} c_{\bm{k}s\sigma} - \Delta_{ss';\eta}^{\sigma\sigma'}g_{\eta}(\bm{k}) \langle c_{\bm{k}s\sigma}^\dagger c_{-\bm{k}s'\sigma'}^\dagger \rangle \Big],
\end{equation}
where the pairing amplitudes \(\Delta_{ss';\eta}^{\sigma\sigma'}\) are determined by the self-consistent equations:
\begin{equation}
\label{eq:Delta_general}
    \Delta_{ss';\eta}^{\sigma\sigma'} = -\frac{V_{ss';\eta}^{\sigma\sigma'} }{N}\sum_{\bm{k'}} g_{\eta}^*(\bm{k}') \langle c_{-\bm{k}'s'\sigma'} c_{\bm{k}'s\sigma} \rangle.
\end{equation}
These self-consistent equations correspond to those in Eqs.~\eqref{eq:self_cons_eq_singlet} and \eqref{eq:self_cons_eq_triplet} of the main text when the interaction coefficients \(V_{ss';\eta}^{\sigma\sigma'}\) are substituted into these equations for each harmonic component \(\eta\).

Using the relationship in Eq.~\eqref{eq:Delta_general}, the last term in Eq.~\eqref{eq:H_p_4} can be expressed as
\begin{equation}
\begin{aligned}
\label{eq:pairing_en_cost}
    & \sum_{s,s'}\sum_{\sigma, \sigma'} \sum_{\eta} \Delta_{ss';\eta}^{\sigma\sigma'} \bigg[-\sum_{\bm{k}}g_{\eta}(\bm{k}) \langle c_{\bm{k}s\sigma}^\dagger c_{-\bm{k}s'\sigma'}^\dagger \rangle\bigg]  \\
    = & \; \sum_{s,s'}\sum_{\sigma, \sigma'} \sum_{\eta} \Delta_{ss';\eta}^{\sigma\sigma'} \bigg[-\sum_{\bm{k}}g_{\eta}(\bm{k}) \langle c_{-\bm{k}'s'\sigma'} c_{\bm{k}'s\sigma} \rangle \bigg]^*  \\
    = & \; N\sum_{s,s'}\sum_{\sigma, \sigma'} \sum_{\eta} \frac{|\Delta_{ss';\eta}^{\sigma\sigma'}|^2}{V_{ss';\eta}^{\sigma\sigma'}}.
\end{aligned}
\end{equation}
As a result, we obtain the pairing Hamiltonian in the general form:
\begin{equation}
\label{eq:H_p_5}
    \hat{H}_\textrm{p} = \sum_{\bm{k}} \sum_{s,s'} \sum_{\sigma,\sigma'} \sum_{\eta} \bigg[ \Delta_{ss';\eta}^{\sigma\sigma'} g_{\eta}(\bm{k})\, c_{\bm{k}s\sigma}^\dagger c_{-\bm{k}s'\sigma'}^\dagger + \left[\Delta_{ss';\eta}^{\sigma\sigma'} g_{\eta}(\bm{k})\right]^* c_{-\bm{k}s'\sigma'} c_{\bm{k}s\sigma} \bigg] + N\sum_{s,s'} \sum_{\sigma,\sigma'} \sum_{\eta} \frac{|\Delta_{ss';\eta}^{\sigma\sigma'}|^2}{V_{ss';\eta}^{\sigma\sigma'}}.
\end{equation}
To derive the spin-singlet BdG Hamiltonian in Eq.~\eqref{eq:H_BdG_singlet} from this general form, we keep \(\eta = \{s, es, d+id, d-id\}\) in the sum and use the pairing gap function definition in Eq.~\eqref{eq:Delta_k_singlet} of the main text. Similarly, the spin-triplet BdG Hamiltonian in Eq.~\eqref{eq:H_BdG_triplet} is derived by taking \(\eta = \{f, p+ip, p-ip, p_z\}\) are used in the sum and using the pairing gap function defined in Eq.~\eqref{eq:Delta_k_triplet}. The derivation of the explicit expression for the pairing energy cost in the last term for both spin-singlet and spin-triplet states is presented in Appendices \ref{app:BdG_analysis_singlet} and \ref{app:BdG_analysis_triplet}, respectively. It is worth noting that the form factors for the interlayer pairing \(\eta = \{s_z, p_z\}\) can support both spin-singlet and spin-triplet states, because they involve different sublattices. However, for simplicity, we focus on the spin-singlet state associated with \(s_z\) and the spin-triplet state associated with \(p_z\).

\section{BdG analysis of spin-singlet states} \label{app:BdG_analysis_singlet}

In this section, we detail the methods for solving the self-consistent gap equations for spin-singlet states and calculating the condensation energy of these states.

\subsection{Derivation of self-consistent gap equations}

We begin by deriving the explicit forms of the self-consistent gap equations. The BdG Hamiltonian for spin-singlet states is given by:
\begin{equation}\label{eq:H_BdG_singlet_app}
    \hat{H}_{BdG} = \sum_{\bm{k}} \Psi_{\bm{k}}^\dagger \mathcal{H}(\bm{k}) \Psi_{\bm{k}} + 2 \sum_{\bm{k}} \epsilon_{0}(\bm{k}) + E_\Delta, 
\end{equation}
where the Nambu spinor is given by $\Psi_{\bm{k}}= ( c_{\bm{k}a\uparrow}, c_{\bm{k}b\uparrow}, c_{-\bm{k}a\downarrow}^\dagger, c_{-\bm{k}b\downarrow}^\dagger)^T$ and the matrix $\mathcal{H}(\bm{k})$ is expressed as:
\begin{equation}
    \mathcal{H}(\bm{k}) = \begin{pmatrix} T_{aa}^{\uparrow\uparrow}(\bm{k}) & T_{ab}^{\uparrow\uparrow}(\bm{k}) & \Delta_{aa}^{\uparrow\downarrow}(\bm{k}) & \Delta_{ab}^{\uparrow\downarrow}(\bm{k})  \\
    T_{ba}^{\uparrow\uparrow}(\bm{k}) & T_{bb}^{\uparrow\uparrow}(\bm{k}) & \Delta_{ba}^{\uparrow\downarrow}(\bm{k}) & \Delta_{bb}^{\uparrow\downarrow}(\bm{k}) \\
    \Delta_{aa}^{\uparrow\downarrow}(\bm{k})^* & \Delta_{ba}^{\uparrow\downarrow}(\bm{k})^* & -T_{aa}^{\downarrow\downarrow}(\bm{k}) & - T_{ab}^{\downarrow\downarrow}(\bm{k}) \\
    \Delta_{ab}^{\uparrow\downarrow}(\bm{k})^* & \Delta_{bb}^{\uparrow\downarrow}(\bm{k})^* & - T_{ba}^{\downarrow\downarrow}(\bm{k}) & -T_{bb}^{\downarrow\downarrow}(\bm{k}) \end{pmatrix}. \label{eq:Hk_singlet}
\end{equation}
The four pairing gap functions $\Delta_{aa}^{\uparrow\downarrow}(\bm{k})$, $\Delta_{bb}^{\uparrow\downarrow}(\bm{k})$, $\Delta_{ab}^{\uparrow\downarrow}(\bm{k})$, and $\Delta_{ba}^{\uparrow\downarrow}(\bm{k})$ are given by:
\begin{equation}
\begin{aligned}
    \Delta_{aa}^{\uparrow\downarrow}(\bm{k}) & = 2\Delta_{aa;s}^{\uparrow\downarrow} + 2g_{es}(\bm{k})\Delta_{aa;es}^{\uparrow\downarrow} + 2g_{d+id}(\bm{k}) \Delta_{aa;d+id}^{\uparrow\downarrow} + 2g_{d-id}(\bm{k}) \Delta_{aa;d-id}^{\uparrow\downarrow}, \\
    \Delta_{bb}^{\uparrow\downarrow}(\bm{k}) & = 2\Delta_{bb;s}^{\uparrow\downarrow} + 2g_{es}(\bm{k})\Delta_{bb;es}^{\uparrow\downarrow} + 2g_{d+id}(\bm{k}) \Delta_{bb;d+id}^{\uparrow\downarrow} + 2g_{d-id}(\bm{k}) \Delta_{bb;d-id}^{\uparrow\downarrow}, \\
    \Delta_{ab}^{\uparrow\downarrow}(\bm{k}) & = 2g_{s_z}(\bm{k})\Delta_{ab;s_z}^{\uparrow\downarrow}, \\
    \Delta_{ba}^{\uparrow\downarrow}(\bm{k}) & = 2g_{s_z}(\bm{k})\Delta_{ba;s_z}^{\uparrow\downarrow}.
\end{aligned}
\end{equation}
Here, the factors of 2 stem from the same contribution from two symmetry-related components in each harmonic component:
\begin{equation}
    \begin{aligned}
     \label{eq:Delta_identities_singlet}
        & \; \Delta_{aa;\uparrow\downarrow}^{s} = -\Delta_{aa;\downarrow\uparrow}^{s}, ~ \Delta_{bb;\uparrow\downarrow}^{s} = - \Delta_{bb;\downarrow\uparrow}^{s}, \\
        & \; \Delta_{aa;\uparrow\downarrow}^{es} = -\Delta_{aa;\downarrow\uparrow}^{es}, ~ \Delta_{bb;\uparrow\downarrow}^{es} = - \Delta_{bb;\downarrow\uparrow}^{es}, \\
        & \; \Delta_{aa;\uparrow\downarrow}^{d\pm id} = -\Delta_{aa;\downarrow\uparrow}^{d\pm id}, ~ \Delta_{bb;\uparrow\downarrow}^{d\pm id} = -\Delta_{bb;\downarrow\uparrow}^{d\pm id}, \\
        & \; \Delta_{ab;\uparrow\downarrow}^{s_z} = -\Delta_{ba;\downarrow\uparrow}^{s_z}, ~ \Delta_{ba;\uparrow\downarrow}^{s_z} = - \Delta_{ab;\downarrow\uparrow}^{s_z}.
        \end{aligned}
\end{equation}
Through the Bogoliubov transformation:
\begin{equation}
\label{eq:BdG_tr_singlet}
    \Psi_{\bm{k}} = U^\dagger(\bm{k}) \tilde{\Psi}_{\bm{k}},
\end{equation}
where $\tilde{\Psi}_{\bm{k}} = (\gamma_{\bm{k}1}, \gamma_{\bm{k}2}, \gamma_{-\bm{k}3}^\dagger, \gamma_{-\bm{k}4}^\dagger)^T$, the BdG Hamiltonian can be diagonalized as:
\begin{equation}
    \hat{H}_{BdG} = \sum_{l=1}^4 \sum_{\bm{k}} E_{\bm{k}l}\gamma_{\bm{k}l}^\dagger \gamma_{\bm{k}l} + 2\sum_{\bm{k}}\epsilon_{0}(\bm{k}) - \sum_{l=3}^4 \sum_{\bm{k}}E_{\bm{k}l} + E_\Delta.
\end{equation}
Here, the operator $\gamma_{\bm{k}l}$ creates a BdG quasiparticle with momentum $\bm{k}$ in the $l$-th energy band. The unitary matrix $U^\dagger(\bm{k})$ has the structure:
\begin{equation}
    U^\dagger(\bm{k}) = \begin{pmatrix} u_{\bm{k};a\uparrow,1} & u_{\bm{k};a\uparrow,2} & -v_{\bm{k};a\uparrow,3}^* & -v_{\bm{k};a\uparrow,4}^* \\ u_{\bm{k};b\uparrow,1} & u_{\bm{k};b\uparrow,2} & -v_{\bm{k};b\uparrow,3}^* & -v_{\bm{k};b\uparrow,4}^* \\ 
    v_{\bm{k};a\downarrow,1} & v_{\bm{k};a\downarrow,2} & u_{\bm{k};a\downarrow,3}^* & u_{\bm{k};a\downarrow,4}^* \\ 
    v_{\bm{k};b\downarrow,1} & v_{\bm{k};b\downarrow,2} & u_{\bm{k};b\downarrow,3}^* & u_{\bm{k};b\downarrow,4}^* \end{pmatrix}, \label{eq:Uk_s0}
\end{equation}
where the Bogoliubov amplitudes $u_{\bm{k};s\sigma,l}$ and $v_{\bm{k};s\sigma,l}$ can be determined by solving the eigenvalue equation:
\begin{equation}
    U(\bm{k}) \mathcal{H}(\bm{k}) U^\dagger(\bm{k}) = \textrm{diag.}\big(E_{\bm{k}1}, E_{\bm{k}2}, -E_{-\bm{k}3}, -E_{-\bm{k}4}\big). \label{eq:UHU_s0}
\end{equation}

The pairing correlation function $\langle c_{-\bm{k}s_1\downarrow} c_{\bm{k}s_2\uparrow} \rangle$ is evaluated in terms of the Bogoliubov amplitudes at zero temperature as:
\begin{equation}
\begin{aligned}
    \langle c_{-\bm{k}s_1\downarrow} c_{\bm{k}s_2\uparrow} \rangle = & \; v_{\bm{k};s_1\downarrow,1}^* u_{\bm{k};s_2\uparrow,1} \Theta(-E_{\bm{k}1}) + v_{\bm{k};s_1\downarrow,2}^* u_{\bm{k};s_2\uparrow,2} \Theta(-E_{\bm{k}2}) \\
    & - u_{\bm{k};s_1\downarrow,3} v_{\bm{k};s_2\uparrow,3}^* [1-\Theta(-E_{-\bm{k}3})] - u_{\bm{k};s_1\downarrow,4} v_{\bm{k};s_2\uparrow,4}^*  [1-\Theta(-E_{-\bm{k}4})]. \label{eq:correl_func_singlet}
\end{aligned}
\end{equation}
Substituting this expression into Eq.~\eqref{eq:Delta_general} yields the self-consistent gap equations used in our mean-field analysis of spin-singlet states:
\begin{equation}
\begin{aligned}
\label{eq:self_cons_eq_singlet}
    \Delta_{ss;s}^{\uparrow\downarrow} = & \; \frac{U}{2N}\sum_{\bm{k}} g_{s}^*(\bm{k}) \Big[u_{\bm{k};s\downarrow,3} v_{\bm{k};s\uparrow,3}^*+ u_{\bm{k};s\downarrow,4}v_{\bm{k};s\uparrow,4}^* \\
    & -v_{\bm{k};s\downarrow,1}^* u_{\bm{k};s\uparrow,1} \Theta(-E_{\bm{k}1}) - v_{\bm{k};s\downarrow,2}^* u_{\bm{k};s\uparrow,2} \Theta(-E_{\bm{k}2}) \\
    & - u_{\bm{k};s\downarrow,3} v_{\bm{k};s\uparrow,3}^* \Theta(-E_{\bm{k}3}) - u_{\bm{k};s\downarrow,4} v_{\bm{k};s\uparrow,4}^* \Theta(-E_{\bm{k}4}) \Big], \\
    \Delta_{ss;es}^{\uparrow\downarrow} = & \; \frac{V_{1}}{3N}\sum_{\bm{k}} g_{es}^*(\bm{k}) \Big[u_{\bm{k};s\downarrow,3} v_{\bm{k};s\uparrow,3}^*+ u_{\bm{k};s\downarrow,4}v_{\bm{k};s\uparrow,4}^* \\
    & -v_{\bm{k};s\downarrow,1}^* u_{\bm{k};s\uparrow,1} \Theta(-E_{\bm{k}1}) - v_{\bm{k};s\downarrow,2}^* u_{\bm{k};s\uparrow,2} \Theta(-E_{\bm{k}2}) \\
    & - u_{\bm{k};s\downarrow,3} v_{\bm{k};s\uparrow,3}^* \Theta(-E_{\bm{k}3}) - u_{\bm{k};s\downarrow,4} v_{\bm{k};s\uparrow,4}^* \Theta(-E_{\bm{k}4}) \Big], \\
    \Delta_{ss;d\pm id}^{\uparrow\downarrow} = & \; \frac{V_{1}}{3N}\sum_{\bm{k}} g_{d\pm id}^*(\bm{k}) \Big[u_{\bm{k};s\downarrow,3} v_{\bm{k};s\uparrow,3}^*+ u_{\bm{k};s\downarrow,4}v_{\bm{k};s\uparrow,4}^* \\
    & -v_{\bm{k};s\downarrow,1}^* u_{\bm{k};s\uparrow,1} \Theta(-E_{\bm{k}1}) - v_{\bm{k};s\downarrow,2}^* u_{\bm{k};s\uparrow,2} \Theta(-E_{\bm{k}2}) \\
    & - u_{\bm{k};s\downarrow,3} v_{\bm{k};s\uparrow,3}^* \Theta(-E_{\bm{k}3}) - u_{\bm{k};s\downarrow,4} v_{\bm{k};s\uparrow,4}^* \Theta(-E_{\bm{k}4}) \Big], \\
    \Delta_{s\bar{s};s_{z}}^{\uparrow\downarrow} = & \; \frac{V_{2}}{2N}\sum_{\bm{k}} g_{s_{z}}^*(\bm{k}) \Big[u_{\bm{k};\bar{s}\downarrow,3} v_{\bm{k};s\uparrow,3}^*+ u_{\bm{k};\bar{s}\downarrow,4}v_{\bm{k};s\uparrow,4}^* \\
    & -v_{\bm{k};\bar{s}\downarrow,1}^* u_{\bm{k};s\uparrow,1} \Theta(-E_{\bm{k}1}) - v_{\bm{k};\bar{s}\downarrow,2}^* u_{\bm{k};s\uparrow,2} \Theta(-E_{\bm{k}2}) \\
    & - u_{\bm{k};\bar{s}\downarrow,3} v_{\bm{k};s\uparrow,3}^* \Theta(-E_{\bm{k}3}) - u_{\bm{k};\bar{s}\downarrow,4} v_{\bm{k};s\uparrow,4}^* \Theta(-E_{\bm{k}4}) \Big].
\end{aligned}
\end{equation}

\subsection{Methods for solving self-consistent gap equations}

We solve the self-consistent gap equations using an iterative method. Initially, the pairing amplitudes are set as follows: for the fully-mixed state, \(\Delta_{ss;s}^{\uparrow\downarrow} = \Delta_{ss;es}^{\uparrow\downarrow} = \Delta_{ss;d+id}^{\uparrow\downarrow} = \Delta_{s\bar{s};s_z}^{\uparrow\downarrow} = 0.3\); for other mixed or pure states, the corresponding relevant pairing amplitudes are initialized at the same value, with all other irrelevant amplitudes set to zero. Next, the Bogoliubov amplitudes and BdG quasiparticle energies are obtained by diagonalizing the BdG Hamiltonian constructed from the current pairing amplitudes. These results are then used to recalculate the pairing amplitudes via the gap equations. This cycle—computing the Bogoliubov amplitudes and energies, and updating the pairing amplitudes—is repeated for 40 iterations, which is sufficient to ensure numerical convergence. All solutions, whether for mixed or pure states, are verified by substituting back into the gap equations.

Using the pairing amplitudes derived from the self-consistent gap equations, we calculate the condensation energy of spin-singlet states with the following expression:
\begin{equation}
\begin{aligned}
\label{eq:condens_energy_singlet}
    F_0 = & \; 2\sum_{\bm{k}}\epsilon_{0}(\bm{k}) - \sum_{l=3}^4 \sum_{\bm{k}} E_{\bm{k}l} + E_\Delta + \sum_{l=1}^4 \sum_{\bm{k}}\Theta(-E_{\bm{k}l})E_{\bm{k}l} \\
    & - \sum_{\alpha=\pm}\sum_{\sigma} \sum_{\bm{k}}\Theta(-E_{\alpha\beta}(\bm{k})) E_{\alpha\beta}(\bm{k}).
\end{aligned}
\end{equation}
This expression represents the energy difference between the superconducting ground state and the normal state. The superconducting ground state energy, corresponding to the BdG state with no negative-energy quasiparticles, is given by:
\begin{equation}
\label{eq:ground_state_energy_singlet}
    E_\textrm{sc} = 2\sum_{\bm{k}}\epsilon_{0}(\bm{k}) - \sum_{l=3}^4 \sum_{\bm{k}} E_{\bm{k}l} + E_\Delta + \sum_{l=1}^4 \sum_{\bm{k}}\Theta(-E_{\bm{k}l})E_{\bm{k}l}.
\end{equation}
The normal state energy, where all pairing amplitudes are zero, is expressed as:
\begin{equation}
    E_\textrm{ns} = \sum_{\alpha=\pm}\sum_{\sigma} \sum_{\bm{k}}\Theta(-E_{\alpha\beta}(\bm{k})) E_{\alpha\beta}(\bm{k}), \label{eq:normal_state_energy}
\end{equation}
where \(\Theta(x)\) is the Heaviside step function. By using the relationships in Eqs.~\eqref{eq:V_coeff} and \eqref{eq:Delta_identities_singlet}, the pairing energy cost $E_\Delta$ is expressed as:
\begin{equation}
    E_\Delta = N \sum_{s} \Bigg(\frac{4\lvert\Delta_{ss;s}^{\uparrow\downarrow}\rvert^2}{U} + \frac{6|\Delta_{ss;es}^{\uparrow\downarrow}|^2}{V_1} + \frac{6|\Delta_{ss;d+id}^{\uparrow\downarrow}|^2}{V_1} + \frac{6|\Delta_{ss;d-id}^{\uparrow\downarrow}|^2}{V_1} + \frac{4|\Delta_{s\bar{s};s_{z}}^{\uparrow\downarrow}|^2}{V_2} \Bigg). \label{eq:E_Delta_singlet}
\end{equation}

\subsection{BdG quasiparticle energy dispersions}

We derive analytic expressions for the BdG quasiparticle energy spectra in the case where \(\Delta_{ab;s_z}^{\uparrow\downarrow} = 0\). For simplicity, we rewrite the matrix \(\mathcal{H}(\bm{k})\) presented in Eq.~\eqref{eq:Hk_singlet} in a simplified form:
\begin{equation}
\mathcal{H} = \begin{pmatrix}
\epsilon_0 \tau_0 + t_x \tau_x + t_z \tau_z + J \tau_z & \Delta \tau_0 \\
\Delta^* \tau_0 & -\epsilon_0 \tau_0 - t_x \tau_x - t_z \tau_z + J \tau_z
\end{pmatrix}, \label{eq:H_BdG_simple_form_triplet}
\end{equation}
where \(\Delta = \Delta_{ss}^{\uparrow\downarrow}(\bm{k})\), with the momentum dependence \(\bm{k}\) suppressed for notation simplicity. The eigenvalue equation associated with this matrix reads:
\begin{equation}
\begin{pmatrix}\epsilon_0 \tau_0+t_{x}\tau_x+t_{z}\tau_{z}+J\tau_{z}&\Delta\tau_0\\\Delta^*\tau_0&-\epsilon_0 \tau_0 -t_{x}\tau_x-t_{z}\tau_{z}+J\tau_{z}\end{pmatrix}\begin{pmatrix}\chi\\\psi\end{pmatrix}=\lambda\begin{pmatrix}\chi\\\psi\end{pmatrix},
\end{equation} 
where \(\alpha = \pm\) and \(\beta = \pm\) label the four solutions. Solving for \(\chi\), we obtain:
\begin{equation}
\Big[\lambda^{2}-\epsilon_0 ^{2}-|\Delta|^{2}-t_{z}^{2}-t_{x}^{2}+J^{2}\Big] \chi = 2 \Big[\epsilon_0 t_{x}\tau_x + (\epsilon_0 t_{z}+\lambda J)\tau_{z} - iJt_{x} \tau_{y}\Big] \chi. \label{eq:equation_wrt_chi_singlet}
\end{equation}
This indicates that \(\chi\) is an eigenvector of the operator \(\epsilon_0 t_x \tau_x + (\epsilon_0 t_z + \lambda J) \tau_z - i J t_x \tau_y\). The eigenvector \(\chi\) can be expressed as:
\begin{equation}
\chi = \begin{pmatrix} \rho + \epsilon_0 t_{z}+\rho J \\ (\epsilon_0 + J)t_{x} \end{pmatrix}, \label{eq:chi}
\end{equation}
and eigenvalues \(\rho \) are:
\begin{equation}
\rho = \pm \sqrt{(\epsilon_0 t_{z} + \lambda J)^{2}+(\epsilon_0^{2}-J^{2})t_{x}^{2}}. \label{eq:lambda_singlet}
\end{equation}
With Eq. \eqref{eq:lambda_singlet}, the eigenvalue equation in Eq. \eqref{eq:equation_wrt_chi_singlet} can be written as
\begin{equation}
\begin{aligned}
\big[\lambda^{2}-(\epsilon_0 ^{2}+t_{x}^{2}+t_{z}^{2}+J^{2}+|\Delta|^{2})\big]^{2}-4\big[\epsilon_0 ^{2}(t_{x}^{2}+t_{z}^{2}+J^{2})+J^{2}(|\Delta|^{2}+t_{z}^{2})\big]-8\lambda\epsilon_0 t_{z}J=0. \label{eq:quartic_equation_singlet} 
\end{aligned} 
\end{equation}
To solve this quartic equation, we apply Ferrari's method. Rearranging Eq.~\eqref{eq:quartic_equation_singlet}, we write:
\begin{equation}
\big[\lambda^{2}-(\epsilon_0 ^{2}+t_{x}^{2}+t_{z}^{2}+J^{2}+|\Delta|^{2})+2S\big]^{2}=\left(2\sqrt{S}\lambda+\frac{2\epsilon_0 t_{z}J}{\sqrt{S}}\right)^{2},\label{eq:biquadratic_equation_singlet}
\end{equation} 
which reduces to two quadratic equations:
\begin{equation} 
\lambda^{2}-(\epsilon_0 ^{2}+t_{x}^{2}+t_{z}^{2}+J^{2}+|\Delta|^{2})+2S=2m\sqrt{S}\lambda+m\frac{2\epsilon_0 t_{z}J}{\sqrt{S}}, \label{eq:two_quadratic_equations_singlet}
\end{equation} 
where $m=\pm$. Solving these, the eigenvalues are given by:
\begin{equation} 
\lambda = m \sqrt{S} + n\sqrt{\epsilon_0^{2}+t_{x}^{2}+t_{z}^{2}+J^{2}+|\Delta|^{2}+m\frac{2\epsilon_0 t_{z}J}{\sqrt{S}}-S},
\end{equation} 
where $n=\pm$. The parameter \(S\) is determined from the following equation:
\begin{equation} 
S^{2}-S(\epsilon_0 ^{2}+t_{x}^{2}+t_{z}^{2}+J^{2}+|\Delta|^{2})+\epsilon_0 ^{2}(t_{x}^{2}+t_{z}^{2}+J^{2})+J^{2}(|\Delta|^{2}+t_{z}^{2})=\frac{\epsilon_0 ^{2}t_{z}^{2}J^{2}}{S}.
\end{equation} 
The solution to this cubic equation can be obtained using Cardano's formula:
\begin{equation}
S=\frac{1}{3}(\epsilon_0 ^{2}+t_{x}^{2}+t_{z}^{2}+J^{2}+|\Delta|^{2}) + \frac{-1 - i\sqrt{3}}{2} \sqrt[3]{-q+\sqrt{q^{2}+p^{3}}}+ \frac{-1 + i\sqrt{3}}{2} \sqrt[3]{-q-\sqrt{q^{2}+p^{3}}},
\end{equation} 
where the parameters $q$ and $p$ are given by
\begin{equation}
\begin{aligned}
q&=-\frac{1}{27}(\epsilon_0 ^{2}+t_{x}^{2}+t_{z}^{2}+J^{2}+|\Delta|^{2})^{3}+\frac{1}{6}(\epsilon_0 ^{2}+t_{x}^{2}+t_{z}^{2}+J^{2}+|\Delta|^{2})\big[\epsilon_0 ^{2}(t_{x}^{2}+t_{z}^{2}+J^{2})+J^{2}(|\Delta|^{2}+t_{z}^{2})\big]-\frac{1}{2}\epsilon_0 ^{2}t_{z}^{2}J^{2}, \\
p&=\frac{1}{3}\big[\epsilon_0 ^{2}(t_{x}^{2}+t_{z}^{2}+J^{2})+J^{2}(|\Delta|^{2}+t_{z}^{2})\big]-\frac{1}{9}(\epsilon_0 ^{2}+t_{x}^{2}+t_{z}^{2}+J^{2}+|\Delta|^{2})^{2}.
\end{aligned}
\end{equation}

At small \(J\), we find that the two eigenvalues for \(n=+\) are positive, while those for \(n=-\) are negative. The signs of these latter two eigenvalues should be flipped, as they correspond to quasiparticle operators in the flipped order: \(\gamma_{-\bm{k}l}\gamma_{\bm{k}l}^\dagger\) for \(l=3,4\), as presented in Eq. \eqref{eq:BdG_tr_singlet}. Accounting for this, the four eigenvalues can be arranged to yield the energy dispersions for the four BdG quasiparticles:
\begin{equation}
    \Xi_{\alpha\beta} = \sqrt{\epsilon_0 ^{2}+t_{x}^{2}+t_{z}^{2}+J^{2}+|\Delta|^{2}+\alpha\frac{2\epsilon_0 t_{z}J}{\sqrt{S}}-S} + \beta \sqrt{S},
\end{equation}
which corresponds to \(\Xi_{\alpha\beta}^\textrm{singlet}(\bm{k})\) with the identification \(\Delta = \Delta_{\uparrow\downarrow}(\bm{k})=\Delta_{aa}^{\uparrow\downarrow}(\bm{k})=\Delta_{bb}^{\uparrow\downarrow}(\bm{k})\) as shown in Eq. \eqref{eq:quasiparticle_energy_singlet} of the main text.

\section{BdG analysis of spin-triplet states} \label{app:BdG_analysis_triplet}

\subsection{Self-consistent gap equations}

In this section, we present the methods for solving the self-consistent gap equations for spin-triplet states and evaluating the condensation energy of spin-triplet states. The BdG Hamiltonian for spin-triplet states is expressed as:
\begin{equation}
\label{eq:H_BdG_triplet_app}
    \hat{H}_{BdG} = \frac{1}{2}\sum_{\sigma}\sum_{\bm{k}} \Psi_{\bm{k}\sigma}^\dagger \mathcal{H}_{\sigma}(\bm{k}) \Psi_{\bm{k}\sigma} + 2 \sum_{\bm{k}} \epsilon_{0}(\bm{k}) + E_\Delta,
\end{equation}
where the Nambu spinor is given by $\Psi_{\bm{k}\sigma}=[c_{\bm{k}a\sigma}, c_{\bm{k}b\sigma}, c_{-\bm{k}a\sigma}^\dagger, c_{-\bm{k}b\sigma}^\dagger ]^T$ and the matrix $\mathcal{H}_{\sigma}(\bm{k})$ is expressed as :
\begin{equation}
    \mathcal{H}_{\sigma}(\bm{k}) = \begin{pmatrix} T_{aa}^{\sigma\sigma}(\bm{k}) & T_{ab}^{\sigma\sigma}(\bm{k}) & \Delta_{aa}^{\sigma\sigma}(\bm{k}) & \Delta_{ab}^{\sigma\sigma}(\bm{k})  \\
    T_{ba}^{\sigma\sigma}(\bm{k}) & T_{bb}^{\sigma\sigma}(\bm{k}) & \Delta_{ba}^{\sigma\sigma}(\bm{k}) & \Delta_{bb}^{\sigma\sigma}(\bm{k}) \\
    \Delta_{aa}^{\sigma\sigma}(\bm{k})^* & \Delta_{ba}^{\sigma\sigma}(\bm{k})^* & - T_{aa}^{\sigma\sigma}(\bm{k}) & - T_{ab}^{\sigma\sigma}(\bm{k}) \\
    \Delta_{ab}^{\sigma\sigma}(\bm{k})^* & \Delta_{bb}^{\sigma\sigma}(\bm{k})^* & - T_{ba}^{\sigma\sigma}(\bm{k}) & - T_{bb}^{\sigma\sigma}(\bm{k}) \end{pmatrix}. \label{eq:Hk_triplet}
\end{equation}
The four gap functions $\Delta_{aa}^{\sigma\sigma}(\bm{k})$, $\Delta_{bb}^{\sigma\sigma}(\bm{k})$, $\Delta_{ab}^{\sigma\sigma}(\bm{k})$, and $\Delta_{ba}^{\sigma\sigma}(\bm{k})$ are given by:
\begin{equation}
\begin{aligned}
    \Delta_{aa}^{\sigma\sigma}(\bm{k}) & = 2g_{f}(\bm{k})\Delta_{aa;f}^{\sigma\sigma} + 2 g_{p+ip}(\bm{k}) \Delta_{aa;p+ip}^{\sigma\sigma} + 2 g_{p-ip}(\bm{k}) \Delta_{aa;p-ip}^{\sigma\sigma}, \\
    \Delta_{bb}^{\sigma\sigma}(\bm{k}) & = 2g_{f}(\bm{k})\Delta_{bb;f}^{\sigma\sigma} + 2 g_{p+ip}(\bm{k}) \Delta_{bb;p+ip}^{\sigma\sigma} + 2 g_{p-ip}(\bm{k}) \Delta_{bb;p-ip}^{\sigma\sigma}, \\
    \Delta_{ab}^{\sigma\sigma}(\bm{k}) & = 2g_{p_z}(\bm{k})\Delta_{ab;p_z}^{\sigma\sigma}, \\
    \Delta_{ba}^{\sigma\sigma}(\bm{k}) & = 2g_{p_z}(\bm{k})\Delta_{ba;p_z}^{\sigma\sigma}.
\end{aligned}
\end{equation}
Here, the factors of 2 arise from splitting the kinetic energy term $T_{ss'}^{\sigma\sigma}(\bm{k})$ into particle and hole sectors, which introduces the \(\frac{1}{2}\) factor in Eq.~\eqref{eq:Hk_triplet}. Through the Bogoliubov transformation:
\begin{equation}
\label{eq:BdG_tr_triplet}
    \Psi_{\bm{k}\sigma} = U_\sigma^\dagger(\bm{k}) \tilde{\Psi}_{\bm{k}\sigma},
\end{equation}
where $\tilde{\Psi}_{\bm{k}\sigma} = (\gamma_{\bm{k}\sigma1}, \gamma_{\bm{k}\sigma2}, \gamma_{-\bm{k}\sigma1}^\dagger, \gamma_{-\bm{k}\sigma2}^\dagger)^T$, the BdG Hamiltonian can be diagonalized as:
\begin{equation}
    \hat{H}_{BdG} = \sum_{l=1}^2 \sum_{\sigma} \sum_{\bm{k}} E_{\bm{k}\sigma l}\gamma_{\bm{k}\sigma l}^\dagger \gamma_{\bm{k}\sigma l} + 2\sum_{\bm{k}} \epsilon_{0}(\bm{k}) - \frac{1}{2} \sum_{l=1}^2 \sum_{\sigma} \sum_{\bm{k}} E_{\bm{k}\sigma l} + E_\Delta,
\end{equation}
The unitary matrix $U_\sigma^\dagger(\bm{k})$ has the structure:
\begin{equation}
    U_{\sigma}^\dagger (\bm{k})= \begin{pmatrix} u_{\bm{k};a\sigma,1} & u_{\bm{k};a\sigma,2} & -v_{\bm{k};a\sigma,1}^* & -v_{\bm{k};a\sigma,2}^* \\ u_{\bm{k};b\sigma,1} & u_{\bm{k};b\sigma,2} & -v_{\bm{k};b\sigma,1}^* & -v_{\bm{k};b\sigma,2}^* \\ 
    v_{\bm{k};a\sigma,1} & v_{\bm{k};a\sigma,2} & u_{\bm{k};a\sigma,1}^* & u_{\bm{k};a\sigma,2}^* \\ 
    v_{\bm{k};b\sigma,1} & v_{\bm{k};b\sigma,2} & u_{\bm{k};b\sigma,1}^* & u_{\bm{k};b\sigma,2}^* \end{pmatrix} ,
\end{equation}
where the Bogoliubov amplitudes $u_{\bm{k};s\sigma,l}$ and $v_{\bm{k};s\sigma,l}$ can be determined by solving the eigenvalue equation:
\begin{equation}
\label{eq:eigenvalue_problem_triplet}
    U_\sigma(\bm{k}) H_\sigma(\bm{k}) U_\sigma(\bm{k})^\dagger = \textrm{diag.}\big(E_{\bm{k}\sigma1}, E_{\bm{k}\sigma2}, -E_{-\bm{k}\sigma1}, -E_{-\bm{k}\sigma2}\big). 
\end{equation}
It is important to recognize that only two different energy bands arise in contrast to the spin-singlet case.

The pairing correlation function $\langle c_{-\bm{k}s_1\downarrow} c_{\bm{k}s_2\uparrow} \rangle$ is evaluated in terms of the Bogoliubov amplitudes at zero temperature as:
\begin{equation}
\begin{aligned}
    \langle c_{-\bm{k}s_1\sigma} c_{\bm{k}s_2\sigma} \rangle = & \; v_{\bm{k};s_1\sigma,1}^* u_{\bm{k};s_2\sigma,1} \Theta(-E_{\bm{k}\sigma 1}) + v_{\bm{k};s_1\sigma,2}^*u_{\bm{k};s_2\sigma,2} \Theta(-E_{\bm{k}\sigma 2}) \\
    & - u_{\bm{k};s_1\sigma,1} v_{\bm{k};s_2\sigma,1}^* [1-\Theta(-E_{-\bm{k}\sigma 1})] - u_{\bm{k};s_1\sigma,2} v_{\bm{k};s_2\sigma,2}^* [1-\Theta(-E_{-\bm{k}\sigma 2})].
\end{aligned}
\end{equation}
Substituting this expression into Eq.~\eqref{eq:Delta_general}, the gap equations can be found:
\begin{equation}
\begin{aligned}
\label{eq:self_cons_eq_triplet}
    \Delta_{ss;f}^{\sigma\sigma} = & \; \frac{V_{1}}{3N}\sum_{\bm{k}} g_{f}^*(\bm{k}) \Big[u_{\bm{k};s\sigma,1} v_{\bm{k};s\sigma,1}^*+ u_{\bm{k};s\sigma,2}v_{\bm{k};s\sigma,2}^* \\
    & - v_{\bm{k};s\sigma,1}^* u_{\bm{k};s\sigma,1} \Theta(-E_{\bm{k}\sigma 1}) - v_{\bm{k};s\sigma,2}^*u_{\bm{k};s\sigma,2} \Theta(-E_{\bm{k}\sigma 2}) \\
    & - u_{\bm{k};s\sigma,1} v_{\bm{k};s\sigma,1}^* \Theta(-E_{\bm{k}\sigma 1}) - u_{\bm{k};s\sigma,2} v_{\bm{k};s\sigma,2}^* \Theta(-E_{\bm{k}\sigma 2}) \Big], \\
    \Delta_{ss;p\pm ip}^{\sigma\sigma} = & \; \frac{V_{1}}{3N}\sum_{\bm{k}} g_{p\pm ip}^*(\bm{k}) \Big[u_{\bm{k};s\sigma,1} v_{\bm{k};s\sigma,1}^*+ u_{\bm{k};s\sigma,2}v_{\bm{k};s\sigma,2}^* \\
    & - v_{\bm{k};s\sigma,1}^* u_{\bm{k};s\sigma,1} \Theta(-E_{\bm{k}\sigma 1}) - v_{\bm{k};s\sigma,2}^*u_{\bm{k};s\sigma,2} \Theta(-E_{\bm{k}\sigma 2}) \\
    & - u_{\bm{k};s\sigma,1} v_{\bm{k};s\sigma,1}^* \Theta(-E_{\bm{k}\sigma 1}) - u_{\bm{k};s\sigma,2} v_{\bm{k};s\sigma,2}^* \Theta(-E_{\bm{k}\sigma 2}) \Big], \\
    \Delta_{s\bar{s};p_{z}}^{\sigma\sigma} = & \; \frac{V_{2}}{2N}\sum_{\bm{k}} g_{p_{z}}^*(\bm{k}) \Big[u_{\bm{k};\bar{s}\sigma,1} v_{\bm{k};s\sigma,1}^*+ u_{\bm{k};\bar{s}\sigma,2}v_{\bm{k};s\sigma,2}^* \\
    & - v_{\bm{k};\bar{s}\sigma,1}^* u_{\bm{k};s\sigma,1} \Theta(-E_{\bm{k}\sigma 1}) - v_{\bm{k};\bar{s}\sigma,2}^*u_{\bm{k};s\sigma,2} \Theta(-E_{\bm{k}\sigma 2}) \\
    & - u_{\bm{k};\bar{s}\sigma,1} v_{\bm{k};s\sigma,1}^* \Theta(-E_{\bm{k}\sigma 1}) - u_{\bm{k};\bar{s}\sigma,2} v_{\bm{k};s\sigma,2}^* \Theta(-E_{\bm{k}\sigma 2}) \Big].
\end{aligned}
\end{equation}

\subsection{Methods for solving self-consistent gap equations}

Similarly to the spin-singlet case, the self-consistent gap equations are solved iteratively. The pairing amplitudes are initially set as follows: \(\Delta_{ss;f}^{\uparrow\downarrow} = \Delta_{ss;p+ip}^{\uparrow\downarrow} = \Delta_{s\bar{s};p_z}^{\uparrow\downarrow} = 0.3\). For other mixed or pure states, the relevant pairing amplitudes are initialized with the same value, while all other amplitudes are set to zero. This cycle—computing the Bogoliubov amplitudes and energies, and updating the pairing amplitudes—is repeated for 40 iterations, which is sufficient to ensure numerical convergence.

Using the pairing amplitudes obtained from the self-consistent gap equations, we compute the condensation energy of spin-triplet states with the following formula:
\begin{equation}
\begin{aligned}
\label{eq:condens_energy_triplet}
    F_0 = & \; 2\sum_{\bm{k}} \epsilon_{0}(\bm{k}) - \frac{1}{2} \sum_{l=1}^2 \sum_{\sigma} E_{\bm{k}\sigma l} + E_\Delta + \sum_{l=1}^2 \sum_{\sigma} \Theta(-E_{\bm{k}\sigma l})E_{\bm{k}\sigma l} \\
    & - \sum_{\alpha=\pm}\sum_{\sigma} \sum_{\bm{k}}\Theta(-E_{\alpha\beta}(\bm{k})) E_{\alpha\beta}(\bm{k}),
\end{aligned}
\end{equation}
which is given by the difference of the ground state energy:
\begin{equation}
\label{eq:ground_state_energy_triplet}
    E_0 = 2\sum_{\bm{k}} \epsilon_{0}(\bm{k}) - \frac{1}{2} \sum_{l=1}^2 \sum_{\sigma} E_{\bm{k}\sigma l} + E_\Delta + \sum_{l=1}^2 \sum_{\sigma} \Theta(-E_{\bm{k}\sigma l})E_{\bm{k}\sigma l},
\end{equation}
and the normal state energy given in Eq.~\eqref{eq:normal_state_energy}. By using the relationship in Eq.~\eqref{eq:V_coeff}, the pairing energy cost $E_\Delta$ is expressed as:
\begin{equation}
\label{eq:E_Delta_triplet}
    E_\Delta = N \sum_{s} \sum_{\sigma} \Bigg(\frac{3|\Delta_{ss;f}^{\sigma\sigma}|^2}{V_1} + \frac{3|\Delta_{ss;p+ip}^{\sigma\sigma}|^2}{V_1} + \frac{2|\Delta_{s\bar{s};p_{z}}^{\sigma\sigma}|^2}{V_2} \Bigg). 
\end{equation}

\subsection{BdG quasiparticle energy dispersions}

We now derive analytic expressions for the BdG quasiparticle energy spectra in the case where \(\Delta_{ab;p_z}^{\sigma\sigma} = 0\). We first consider the spin-up sector \(\mathcal{H}_\uparrow(\bm{k})\). For simplicity, we rewrite the matrix \(\mathcal{H}_\sigma(\bm{k})\) presented in Eq.~\eqref{eq:Hk_triplet} in a simplified form:
\begin{equation}
\mathcal{H} = \begin{pmatrix}
\epsilon_0 \tau_0 + t_x \tau_x + t_z \tau_z + J \tau_z & \Delta_0 \tau_0 + \Delta_z \tau_z \\
 \Delta_0^* \tau_0 + \Delta_z^* \tau_z & -\epsilon_0 \tau_0 - t_x \tau_x - t_z \tau_z - J \tau_z
\end{pmatrix}, \label{eq:H_BdG_simple_form_singlet}
\end{equation}
where \(\Delta_0 = (\Delta_{aa}^{\uparrow\uparrow}(\bm{k})+\Delta_{bb}^{\uparrow\uparrow}(\bm{k}))/2\) and \(\Delta_z = (\Delta_{aa}^{\uparrow\uparrow}(\bm{k})-\Delta_{bb}^{\uparrow\uparrow}(\bm{k}))/2\), with the momentum dependence \(\bm{k}\) suppressed for notation simplicity. We employ a unitary transformation:
\begin{equation}
    W^\dagger = \begin{pmatrix}
        w_1 & -w_2 & 0 & 0 \\
        w_2 & w_1 & 0 & 0 \\
        0 & 0 & w_1 & -w_2 \\
        0 & 0 & w_2 & w_1
    \end{pmatrix},
\end{equation}
where \(w_1\) and \(w_2\) are given by
\begin{equation}
\begin{aligned}
    w_1 & = \sqrt{\frac{1}{2}\Bigg(1+\frac{(t_z+J)}{\sqrt{t_x^2+(t_z+J)^2}}\Bigg)}, \\
    w_2 & = \textrm{sgn}(t_x)\sqrt{\frac{1}{2}\Bigg(1-\frac{(t_z+J)}{\sqrt{t_x^2+(t_z+J)^2}}\Bigg)}, \\
\end{aligned}    
\end{equation}
The sub-block matrix
\(w=\begin{pmatrix}
    w_1 & -w_2 \\
    w_2 & w_1
\end{pmatrix}\) diagonalizes the kinetic part as \(w(\epsilon_0 \tau_0 + t_x \tau_x + t_z \tau_z + J \tau_z)w^\dagger=\epsilon_0 \tau_0 + \sqrt{t_x^2+(t_z+J)^2} \tau_z\). Conducting,
\begin{equation}
    W \mathcal{H} W^\dagger = \begin{pmatrix}
        E_+ & 0 & \Delta_+ & \Delta_c \\
        0 & E_- & \Delta_c & \Delta_- \\
        \Delta_+^* & \Delta_c^* & -E_+ & 0 \\
        \Delta_c^* & \Delta_-^* & 0 & -E_-
    \end{pmatrix},
\end{equation}
where \(E_\pm\), \(\Delta_\pm\), and \(\Delta_c\) are given by
\begin{equation}
\begin{aligned}
    E_\pm = & \; \epsilon_0 \pm \sqrt{t_x^2+(t_z+J)^2}, \\
    \Delta_\pm = & \; \Delta_0 \pm \Delta_z\frac{(t_z+J)}{\sqrt{t_x^2+(t_z+J)^2}}, \\
    \Delta_c = & \; -\Delta_z\frac{t_x}{\sqrt{t_x^2+(t_z+J)^2}}. \label{eq:energy_triplet}
\end{aligned}
\end{equation}
The eigenvalue equation reads
\begin{equation}
\begin{aligned}
    & |W \mathcal{H} W-\lambda I_{4\times 4}| \\
    = & \; \left\lvert\begin{pmatrix}
    E_+-\lambda & 0 & \Delta_+ & \Delta_c \\
    0 & E_--\lambda & \Delta_c & \Delta_- \\
    \Delta_+^* & \Delta_c^* & -E_+-\lambda & 0 \\
    \Delta_c^* & \Delta_-^* & 0 & -E_--\lambda
    \end{pmatrix} \right\rvert \\
    = & \; \lambda^4 - \lambda^2 (E_+^2+E_-^2+|\Delta_+|^2+|\Delta_-|^2+2|\Delta_{c}|^2) + E_+^2E_-^2 + E_+^2|\Delta_-|^2 + E_-^2|\Delta_+|^2 + |\Delta_+\Delta_--\Delta_c^2|^2 + 2|\Delta_c|^2E_+E_-=0.
\end{aligned}
\end{equation}
The four solutions of this equation are given by
\begin{equation}
\begin{aligned}
    \lambda = & \; m \Bigg[\frac{1}{2}\big(E_+^2+|\Delta_+|^2+E_-^2+|\Delta_-|^2+2|\Delta_c|^2\big) + n \Bigg\{\frac{1}{4} \big(E_+^2+|\Delta_+|^2-E_-^2-|\Delta_-|^2\big)^2 \\
    & \; +|\Delta_c|^2\Big((E_+-E_-)^2+|\Delta_+|^2+|\Delta_-|^2\Big)+\Big((\Delta_+\Delta_-)^* \Delta_c^2+(\Delta_+\Delta_-)(\Delta_c^*)^2\Big) \Bigg\}^{\frac{1}{2}} ~ \Bigg]^{\frac{1}{2}},
\end{aligned}
\end{equation}
where \(m=\pm\) and \(n=\pm\). By using that the phases of \(\Delta_+\), \(\Delta_-\), and \(\Delta_c\) are equal, we simplify the above as:
\begin{equation}
\begin{aligned}
    \lambda = & \; m \Bigg[\frac{1}{2}\big(E_+^2+|\Delta_+|^2+E_-^2+|\Delta_-|^2+2|\Delta_c|^2\big) \\
    & + \; n \sqrt{\frac{1}{4} \big(E_+^2+|\Delta_+|^2-E_-^2-|\Delta_-|^2\big)^2 +|\Delta_c|^2\big((E_+-E_-)^2+|\Delta_++\Delta_-|^2\big) } ~ \Bigg]^{\frac{1}{2}},
\end{aligned}
\end{equation}
Using the expressions in Eq. \eqref{eq:energy_triplet}, we simplify 
\begin{equation}
    \lambda = m \bigg[\epsilon_0^2+t_x^2+(t_z+J)^2+|\Delta_0|^2+|\Delta_z|^2 + 2n \sqrt{\big(\epsilon_0(t_z+J)+|\Delta_0||\Delta_z|\big)^2+t_x^2\big(\epsilon_0^2+|\Delta_z|^2\big)} ~ \bigg]^{\frac{1}{2}}.
\end{equation}
For the spin-down sector \(\mathcal{H}_\downarrow(\bm{k})\), the sign in front of \(J\) flips in \(\lambda\). Combining these two cases, we obtain
\begin{equation}
    \lambda = m \bigg[\epsilon_0^2+t_x^2+(t_z+s(\sigma)J)^2+|\Delta_0|^2+|\Delta_z|^2 + 2 n \sqrt{\big(\epsilon_0(t_z+s(\sigma)J)+|\Delta_0||\Delta_z|\big)^2+t_x^2\big(\epsilon_0^2+|\Delta_z|^2\big)} ~ \bigg]^{\frac{1}{2}}.
\end{equation}

For each spin sector, at small \(J\), we find that the two eigenvalues for \(m=+\) are positive, while those for \(m=-\) are negative. The signs of these latter two eigenvalues should be flipped, as they correspond to quasiparticle operators in the flipped order: \(\gamma_{-\bm{k}l}\gamma_{\bm{k}l}^\dagger\) for \(l=1,2\), as presented in Eq. \eqref{eq:BdG_tr_triplet}. Accounting for this, the four eigenvalues can be arranged to yield the energy dispersions for the four BdG quasiparticles:
\begin{equation}
    \Xi_{\alpha\beta} = \bigg[\epsilon_0^2+t_x^2+(t_z+\beta J)^2+|\Delta_0|^2+|\Delta_z|^2 + 2 \alpha \sqrt{\big(\epsilon_0(t_z+\beta J)+|\Delta_0||\Delta_z|\big)^2+t_x^2\big(\epsilon_0^2+|\Delta_z|^2\big)} ~ \bigg]^{\frac{1}{2}}.
\end{equation}
which corresponds to \(\Xi_{\alpha\beta}^\textrm{triplet}(\bm{k})\) as shown in Eq. \eqref{eq:quasiparticle_energy_triplet} of the main text.

\section{Phase diagrams for different interaction parameters} \label{app:phase_diagrams}

\pagebreak

\begin{figure}[t!]
    \centering
    \includegraphics[width=.99\columnwidth]{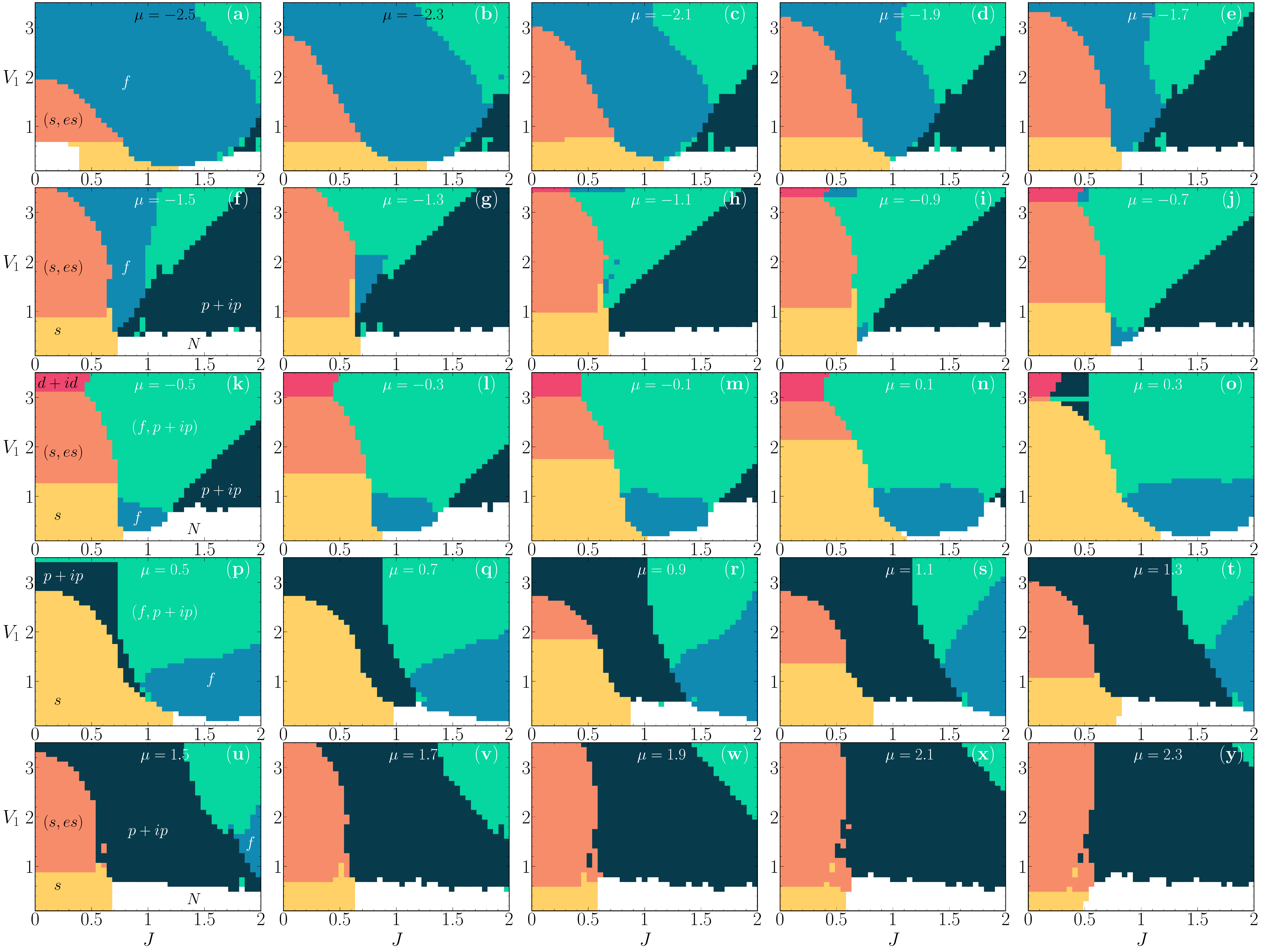}
    \caption{
    \textbf{Zero temperature \(J\)–\(V_1\) phase diagrams for additional \(\mu\) values.} Each panel illustrates the phase diagram for the selected values of \(\mu\) ranging from \(-2.5\) to \(2.3\) in increments of \(0.2\), with the corresponding \(\mu\) value indicated in each panel. The \(x\)-axis represents the \(J\) value, while the \(y\)-axis indicates the \(V_1\) value across all panels. The parameter \(V_2\) is set to \(0\) in all cases.
    }
    \label{fig_s2:phase_diagrams2}
\end{figure}

We demonstrate that the phase diagrams presented in Figs. \ref{fig:phase_diagram}(a)–(c) of the main text remain qualitatively valid across a broad range of \(\mu\) values, as discussed in the second paragraph of Sec. \ref{secIIC}. The \(J\)–\(V_1\) phase diagrams for \(-2.3 \leq \mu \leq -1.3\) [panels (b)–(g) in Fig. \ref{fig_s2:phase_diagrams2}] are consistent with Fig. \ref{fig:phase_diagram}(a), exhibiting the same classes of superconducting phases along with consistent arrangements in the \(J\)–\(V_1\) space. Notable changes from Fig. \ref{fig:phase_diagram}(a) corresponding to \(\mu=-2\) include the expansion of the chiral \(p\)-wave phase and the contraction of the \(f\)-wave phase for \(\mu > -2\), and vice versa for \(\mu < -2\). Similarly, the phase diagrams for \(-0.5 \leq \mu \leq 0.3\) [panels (k)–(o)] are consistent with Fig. \ref{fig:phase_diagram}(b), featuring identical phases and consistent arrangements. Key changes from Fig. \ref{fig:phase_diagram}(b) corresponding to \(\mu=0\) include the expansion of the \(f\)-wave phase and a shift of the chiral \(p\)-wave phase to a lower \(J\) regime for \(\mu < 0\) and vice versa for \(\mu > 0\). Lastly, the phase diagrams for \(1.7 \leq \mu \leq 2.3\) [panels (v)–(y)] remain consistent with Fig. \ref{fig:phase_diagram}(c), displaying identical phases and consistent arrangements. Main changes from Fig. \ref{fig:phase_diagram}(b) include the expansion of the chiral \(p\)-wave phase and a shift of the \(f\)-wave phase to a higher \(J\) regime as \(\mu\) increases. These findings confirm that the phase diagrams presented in Figs. \ref{fig:phase_diagram}(a)–(c) effectively capture three distinct patterns of superconducting phase distribution, as described in the main text.

As briefly discussed in the second paragraph of Sec. \ref{secIIC}, there are two additional patterns distinct from those presented in Figs. \ref{fig:phase_diagram}(a)–(c) within the remaining \(\mu\) ranges. The first pattern appears in the phase diagrams for \(-1.1 \leq \mu \leq -0.7\) [panels (h)–(j)]. These diagrams display the same classes of superconducting phases found in Fig. \ref{fig:phase_diagram}(a), but they also include the chiral \(d\)-wave phase at the highest \(V_1\) range within the spin singlet domain, which is absent in Fig. \ref{fig:phase_diagram}(a). Additionally, the area of the chiral \(p\)-wave phase is significantly larger than in Fig. \ref{fig:phase_diagram}(a), while the area of the \(f\)-wave phase becomes vanishingly small, establishing the chiral \(p\)-wave phase as the dominant triplet phase. The second pattern is observed from the phase diagrams corresponding to \(0.5 \leq \mu \leq 1.1\) [panels (p)–(s)]. In these diagrams, the \((s,es)\)-wave phase is omitted compared to Fig. \ref{fig:phase_diagram}(c). Furthermore, the size of the chiral \(p\)-wave phase gradually decreases as \(\mu\) decreases, becoming significantly smaller than in Fig. \ref{fig:phase_diagram}(c), and it nearly vanishes at \(\mu=0.5\).

\begin{figure}[t!]
    \centering
    \includegraphics[width=.95\columnwidth]{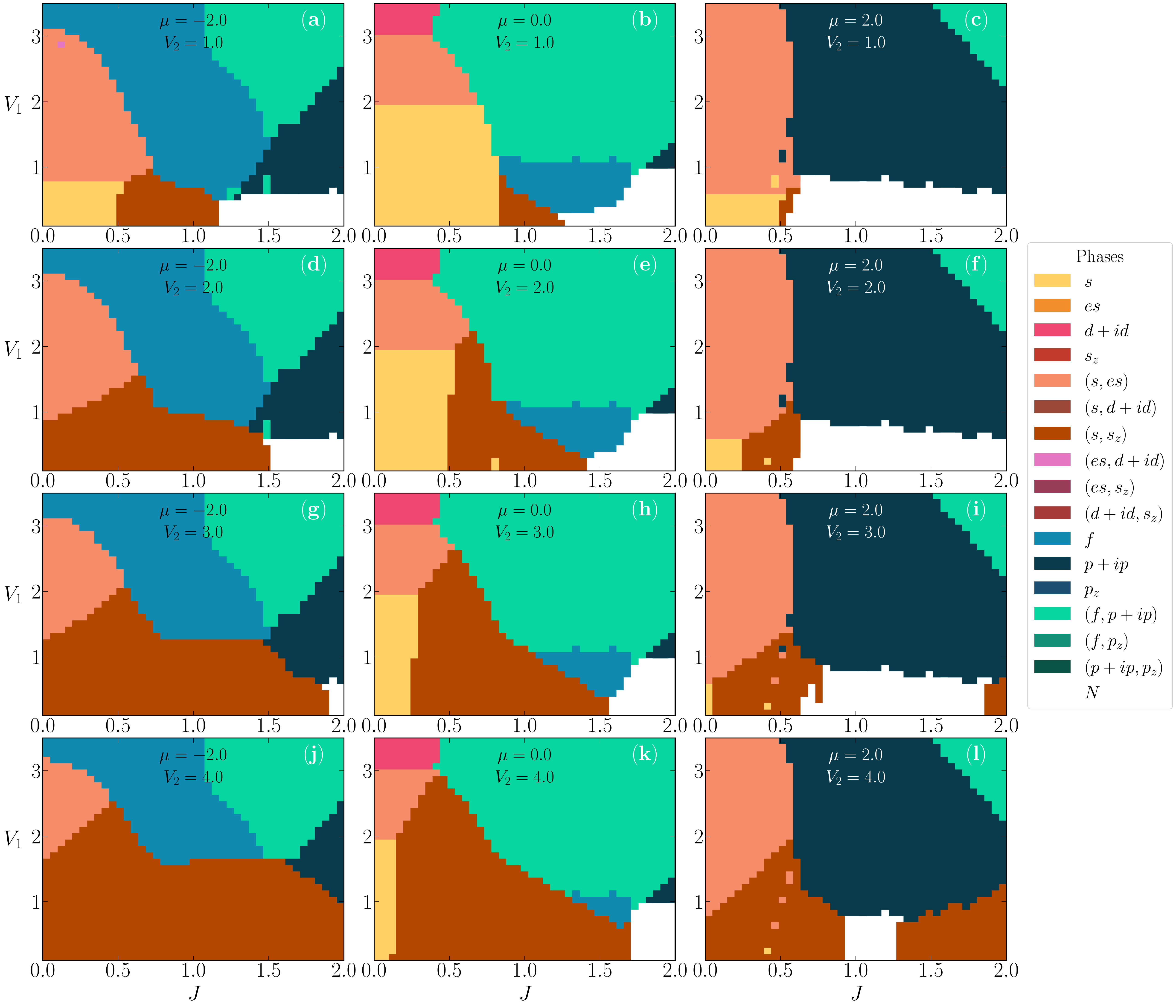}
    \caption{\textbf{Zero temperature \(J\)–\(V_1\) phase diagrams for nonzero \(V_2\) values.} Each panel illustrates the phase diagram for the selected values of \((\mu,V_2)\), as indicated in each panel. The \(x\)-axis represents the \(J\) value, while the \(y\)-axis indicates the \(V_1\) value across all panels. The $(s,s_z)$-wave phase is indicated in brown. The remaining colors follow the same color scheme used to identify the various superconducting phases in Fig.~\ref{fig:phase_diagram}.}
    \label{fig_s4:phase_diagrams4}
\end{figure}

We next demonstrate that the phase diagrams presented in Figs.~\ref{fig:phase_diagram}(a)–(c) of the main text remain qualitatively valid even upon considering the effects of $V_2$, as discussed in Sec.~\ref{secIIC}. Figure~\ref{fig_s4:phase_diagrams4} illustrates the evolution of the $J$–$V_1$ phase diagrams as $V_2$ increases from 1 to 4. For $V_2 \leq 2$, the spin-triplet phases—specifically the $f$-wave, chiral $p$-wave, and $(f,p+ip)$-wave phases—are largely consistent with those observed at $V_2=0$ [see Figs.~\ref{fig_s4:phase_diagrams4}(a)–(f)]. However, for $V_2 > 2$, these triplet phases are partially supplanted by the $(s,s_z)$-wave phase [Figs.~\ref{fig_s4:phase_diagrams4}(g)–(l)]. Notably, while the chiral $p$-wave phase persists strongly at $\mu=2$ [Fig.~\ref{fig_s4:phase_diagrams4}(l)], it is relatively suppressed at $\mu=-2$ [Fig.~\ref{fig_s4:phase_diagrams4}(j)]. Overall, the stability of the spin-triplet phases is attributed to the fact that the $p_z$-wave pairing amplitudes remain consistently zero across all parameter regimes. Furthermore, the region occupied by the chiral $d$-wave phase is unaffected by $V_2$. Consequently, we conclude that the chiral $d$-wave and $p$-wave phases are robust against the influence of $V_2$.

\section{Pair correlation functions} \label{app:pair_correlation_function}

\begin{figure*}[t!]
    \centering    
    \includegraphics[width=\columnwidth]{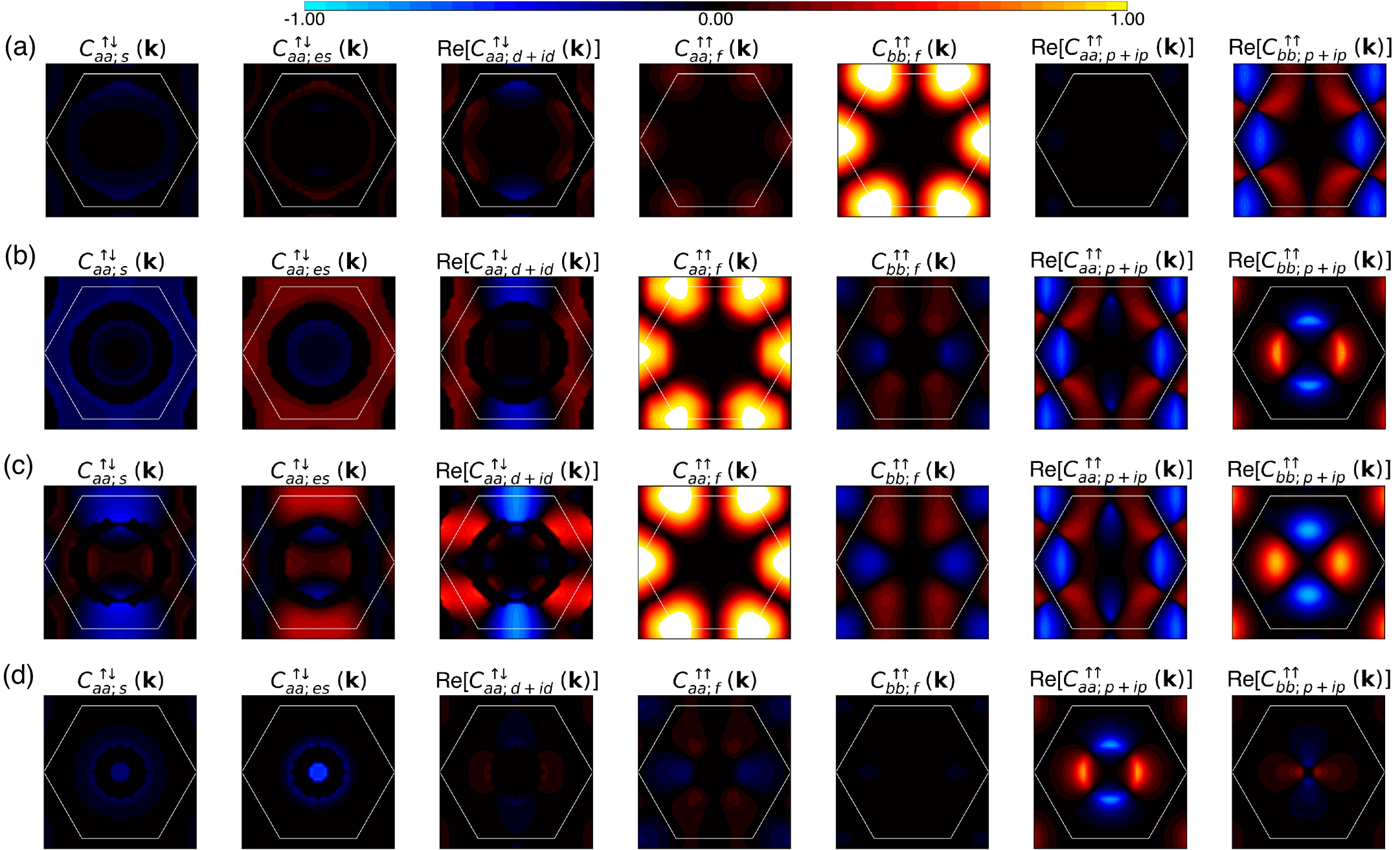}
    \caption{\textbf{Weighted pair correlation functions (WPCFs).} The parameter sets $(\mu, J, V_1)$ for each panel are: (a) $(-2, 0.95, 2.5)$, (b) $(0, 0.85, 2)$, (c) $(0, 0.9, 3.1)$, and (d) $(2, 0.65, 2)$, which correspond to the cases presented in Figs.~\ref{fig:BFS}(a)–(d). The subpanels within each panel are organized as follows: (1st column) $C_{aa; s}^{\uparrow\downarrow}(\bm{k})$, (2nd) $C_{aa; es}^{\uparrow\downarrow}(\bm{k})$, (3rd) $\text{Re}[C_{aa; d+id}^{\uparrow\downarrow}(\bm{k})]$, (4th) $C_{aa; f}^{\uparrow\uparrow}(\bm{k})$, (5th) $C_{bb; f}^{\uparrow\uparrow}(\bm{k})$, (6th) $\text{Re}[C_{aa; p+ip}^{\uparrow\uparrow}(\bm{k})]$, and (7th) $\text{Re}[C_{bb; p+ip}^{\uparrow\uparrow}(\bm{k})]$. In all subpanels, black regions denote BFSs where the WPCF vanishes, and white solid lines indicate the boundaries of the first Brillouin zone. The correlation functions are projected onto the $k_x$--$k_y$ plane at $k_z = \pi$, with both $k_x$ and $k_y$ axes ranging from $-4\pi/3$ to $4\pi/3$. 
    }
    \label{fig:pair_correlation_function}
\end{figure*}

To elucidate how the formation of BFSs suppresses spin-singlet phases and stabilizes spin-triplet phases, we compute the weighted pairing correlation function (WPCF). The WPCF is defined as the pairing correlation function multiplied by the form factor associated with each pairing symmetry:
\begin{equation}
    C_{ss'; \eta}^{\sigma\sigma'}(\bm{k}) = g_{\eta}^*(\bm{k}) \left\langle c_{-\bm{k}s'\sigma'} c_{\bm{k}s\sigma} \right\rangle.
\end{equation}
For spin-singlet pairings, the pairing amplitudes are identical on the $A$ and $B$ sublattices; consequently, the sublattice components $C_{aa}$ and $C_{bb}$ are identical for the $s$-wave, $es$-wave, and chiral $d$-wave channels. For $s$-wave, $es$-wave, and $f$-wave pairings, the imaginary parts vanish, meaning the real parts provide the complete physical description. In the case of chiral $d$-wave and chiral $p$-wave pairings, the imaginary parts are odd functions of $\bm{k}$ that sum to zero upon integration over the BZ. The spin up and down sectors. We therefore omit these imaginary parts to focus on the physically dominant real components. This reduction yields seven representative functions per panel: three spin-singlet WPCFs---$C_{aa; s}^{\uparrow\downarrow}(\bm{k})$, $C_{aa; es}^{\uparrow\downarrow}(\bm{k})$, and $\text{Re}[C_{aa; d+id}^{\uparrow\downarrow}(\bm{k})]$; two $f$-wave WPCFs---$C_{aa; f}^{\uparrow\uparrow}(\bm{k})$ and $C_{bb; f}^{\uparrow\uparrow}(\bm{k})$; and two chiral $p$-wave components---$\text{Re}[C_{aa; p+ip}^{\uparrow\downarrow}(\bm{k})]$ and $\text{Re}[C_{bb; p+ip}^{\uparrow\uparrow}(\bm{k})]$.

Figure~\ref{fig:pair_correlation_function}(a) displays these seven WPCFs at $(\mu, J, V_1)=(-2, 0.95, 2.5)$, corresponding to the BFSs shown in Fig.~\ref{fig:BFS}(a). The BFSs form near the $\Gamma$ point and suppress the spin-singlet WPCFs, thereby explaining the suppression of singlet pairing amplitudes. Conversely, the $f$-wave WPCFs exhibit substantial values near the BZ corners. The chiral $p$-wave WPCFs display both positive and negative values that sum to zero, consistent with the absence of chiral $p$-wave pairing in this regime. Similar behavior is observed for $(\mu, J, V_1)=(0, 0.85, 2)$ in Fig.~\ref{fig:pair_correlation_function}(b), where the BFSs appear as circular rings centered at the $\Gamma$ point, again suppressing the spin-singlet WPCFs. While $C_{aa; f}^{\uparrow\uparrow}(\bm{k})$ remains significant near the BZ corners, leading to a substantial pairing amplitude, $C_{bb; f}^{\uparrow\uparrow}(\bm{k})$ exhibits an opposite sign (odd parity) relative to panel (a). However, these values are not perfectly balanced, resulting in a small but non-zero pairing amplitude. Crucially, the chiral $p$-wave WPCFs show an imbalance between positive and negative values; this net non-zero sum explains the emergence of the mixed $f$-wave and chiral $p$-wave components observed in Fig.~\ref{fig:pairing_amplitudes}(b). A similar trend is found in Fig.~\ref{fig:pair_correlation_function}(c) for $(\mu, J, V_1)=(0, 0.9, 3.1)$, where the chiral $p$-wave components are again unbalanced, stabilizing the mixed phase shown in Fig.~\ref{fig:pairing_amplitudes}(c). Finally, for $(\mu, J, V_1) = (2, 0.7, 2)$ in Fig.~\ref{fig:pair_correlation_function}(d), the spin-singlet components are suppressed by the BFS near the $\Gamma$ point. In this regime, the $f$-wave components are balanced to zero, while the chiral $p$-wave components remain unbalanced, accounting for the dominance of chiral $p$-wave pairing as observed in Fig.~\ref{fig:pairing_amplitudes}(d).

\bibliography{ref}

\end{document}